\documentclass[smallcondensed,numbook,envcountsame,final]{svjour3}
\usepackage[cleanup={aux,dvi,ps,pdf},pspdf={-dALLOWPSTRANSPARENCY -dNOSAFER}]{auto-pst-pdf}
\usepackage[T1]{fontenc}
\usepackage[utf8]{inputenc} 
\usepackage{slantsc}
\usepackage{csquotes}
\usepackage{setspace}
\usepackage{amsmath,amsfonts,amssymb,latexsym,mathtools,bbm,nicefrac}
\usepackage{graphicx,pstricks,pstricks-add}
\usepackage{calc} 
\usepackage{pgf,tikz}
\usepackage{nicefrac}
\usepackage{enumitem}
\usepackage{caption}
\usepackage{multirow}
\usepackage{colortbl,color,esvect}
\usepackage[bb=boondox]{mathalfa}
 \usepackage{mathptmx}
 \usepackage[scaled=.90]{helvet}
 \usepackage{courier}
\smartqed

\usepackage[pagebackref,hyperfootnotes=false,colorlinks=true,urlcolor=blue,linkcolor=blue,citecolor=blue,linktocpage=true]{hyperref}
\renewcommand*{\backref}[1]{}\renewcommand*{\backrefalt}[4]{\ifcase #1 (Not cited.)\or (Cited on page~#2.)\else (Cited on pages~#2.)\fi}
\makeatletter
       \renewcommand*{\@biblabel}[1]{[#1]}
\makeatother

\spnewtheorem*{theorem*}{Theorem}{\normalfont\bfseries}{}
\spnewtheorem*{lemma*}{Lemma}{\normalfont\bfseries}{}
\spnewtheorem*{corollary*}{Corollary}{\normalfont\bfseries}{}
\def\ceq{\coloneqq}
\def\eqc{\eqqcolon}
\newcommand{\V}{\upsilon}
\newcommand{\ini}{\textsf{IN-IN}}
\newcommand{\R}{\mathbbm{R}}

\renewcommand{\P}{\mathbbm{P}}
\newcommand{\B}{\mathcal{B}}

\newcommand{\A}{\mathcal{A}}
\newcommand{\CM}{\textsf{CM}}
\newcommand{\HV}{\textsf{HV}}
\newcommand{\CPT}{\textsf{CPT}}
\newcommand{\QM}{\textsf{QM}}
\newcommand{\dd}{\textrm{d}}
\newcommand{\LL}{\textrm{L}}
\newcommand{\uu}{\mathbbm{1}}
\newcommand{\ii}{\textrm{i}}
\newcommand{\su}{\mathsf{u}}
\newcommand{\sv}{\mathsf{v}}
\newcommand{\kreis}[1]{\unitlength1ex\begin{picture}(2.5,2.5)%
\put(0.75,0.75){\circle{2.5}}\put(0.75,0.75){\makebox(0,0){#1}}\end{picture}}

\newcommand{\one}{~\kreis{1}}
\newcommand{\two}{~\kreis{2}}

\renewcommand{\H}{\mathcal{H}}
\DeclareMathOperator\tr{Tr}
\DeclareMathOperator\re{Re}
\DeclareMathOperator\sgn{sgn}
\DeclareMathOperator\spec{spec}
\DeclareMathOperator\im{Im}
\newcommand{\quotes}[1]{``{}#1''{}}

\DeclareMathOperator\dom{dom}
\newcommand{\C}{\mathbb{C}}

\newcommand{\braket}[1]{\langle #1 \rangle}
\newcommand{\bra}[1]{\left\langle\left. #1 \right|\right.}
\newcommand{\ket}[1]{\left|\left. #1 \right\rangle\right.}

\newcommand{\bordercolor}{red}

\newcommand{\boxedeqn}[1]{\fcolorbox{\bordercolor}{white}{$\displaystyle #1 $}}
\newcommand{\boxedtext}[1]{\fcolorbox{\bordercolor}{white}{#1}}

\usepackage[nodayofweek]{datetime}
\newdateformat{mydate}{\twodigit{\THEDAY}{ }\monthname[\THEMONTH] \THEYEAR}
\mydate

\allowdisplaybreaks

\begin{document}

\phantom
\bigskip\bigskip\bigskip\bigskip\bigskip\bigskip
  \begin{center}
  {\huge The Quantum Formalism Revisited} \bigskip
  
  {\large Differences from the classical theory with some of them quantified}
  
  {\large 100 years after {\scshape Heisenberg}'s discovery}

 \bigskip\bigskip\bigskip\bigskip

{\large Hajo {\scshape Leschke}}

\bigskip
{\small Friedrich--Alexander--Universität Erlangen--Nürnberg}\\
{\small Institut für Theoretische Physik}\\
{\small Erlangen, Germany}


\bigskip\bigskip

To Wolfgang {\scshape Kundt}, who made me understand\\
what he had learned from Pascual {\scshape Jordan} 

\bigskip\bigskip\bigskip\bigskip\bigskip\bigskip\bigskip

Intermediate IQSA 2025 Conference\\
30 June -- 4 July, 2025\\
Tropea, Calabria (Italy)\\

\medskip
 {\small[arXiv:2506.16480]}
 {\small[Date of this version: 30 September 2026]}
 
\bigskip
A slightly different version appeared as the 2nd chapter of the book\\ \textit{Quantum Mechanics: A Century Later}, ed. by Tuck {\scshape Choy}\\
World Scientific, Singapore 2026, ISBN 978-981-98-2756-5

\bigskip\bigskip\bigskip\bigskip\bigskip\bigskip\bigskip\bigskip\bigskip
\textit{\footnotesize
 ``Heisenberg hat ein großes Quantenei gelegt. In Göttingen glauben sie daran (ich nicht).''}\footnote{In English: ``Heisenberg has laid a big quantum egg. In Göttingen they believe it (I don't).''\newline [See also: Tom {\scshape Siegfried} in the magazine \textit {Science News}, March 30, 2020.]}  \\
\smallskip{}
{\scriptsize Albert {\scshape Einstein} to Paul {\scshape Ehrenfest}}\\
{\scriptsize (in a letter dated 20 November 1925)}
\smallskip{}

\centering{------}

\smallskip
\textit {\footnotesize
``If a subject is robust it should be insensitive to its foundations.''}\\
\smallskip
{\scriptsize Mark {\scshape Kac}}\\ 
{\scriptsize (p. 111 in his autobiography \cite{Ka1985})}
\vfill
 \end{center}
\newpage
\begin{abstract}

For the simple system of a point-like particle confined to a straight line, I compile, initially in a concise table, the structural elements of quantum mechanics and contrast them with those of classical (statistical) mechanics. Despite many similarities, there are the well-known fundamental differences, resulting from the algebraic non-commutativity in the quantum structure. The latter was discovered by Werner {\scshape Heisenberg} (1901--1976) in June 1925 on the small island of Helgoland in the North Sea, as a consequence of understanding atomic spectral data within a matrix scheme consistent with energy conservation. I discuss the differences and exemplify their quantifications by the variance and entropic indeterminacy inequalities, by (pseudo-)classical bounds on quantum canonical partition functions, and by the correlation inequalities of John {\scshape Bell} (1928--1990) and others.

\end{abstract}
\bigskip\bigskip\bigskip\bigskip\medskip
\tableofcontents

\mathtoolsset{centercolon}

\bigskip\bigskip\bigskip\bigskip\bigskip\bigskip\bigskip

\section*{Acknowledgments} 
This text grew out of lectures given at the Friedrich--Alexander--Universität Erlangen--Nürn\-berg (FAU). It also owes its existence to a series of lectures for probability theorists 
(Sept. 2013, Villa Volpi, Ghiffa, Piedmont, Italy) suggested by Peter {\scshape Eichelsbacher} (Ruhr-Universi\-tät Bochum (RUB), Germany).
I am especially grateful to my former students Helmut {\scshape Fink} (Heisen\-berg-Gesellschaft, München), Sebastian {\scshape Rothlauf}, and Rainer {\scshape Ruder} for valuable discussions and their help with the \LaTeX\, document-preparation system. I also appreciate many clarifying comments made by Stefan {\scshape Weigert} (University of York, UK) and by my colleague Daniel {\scshape Burgarth} (FAU). Last but not least, my thanks go to the editor Tuck C. {\scshape Choy} for his kind invitation, assistance, and patience.
A travel grant of the Hans-Hermann {\scshape Toedter} Demokritos-Stiftung (München) is gratefully acknowledged.
\newpage

\section{Introduction}
\setlength{\parindent}{1.8em}

\emph {Quantum mechanics} is the most successful theory of physics for model building from subatomic to cosmic length scales. As its name suggests, probably first coined by {\scshape Born} \cite{Bo1924}, the theory was needed because the time-honored and over the centuries developed \emph {classical mechanics} \cite{Kn2018} of {\scshape Newton, Lagrange, Hamilton}, and others had turned out to be unable to explain the \emph{quantized} or portioned energy values observed in atomic physics. Very likely, most physicists had expected a more or less refined version to emerge from classical mechanics. However, after {\scshape Heisenberg}'s groundbreaking publication \cite{Hei1925} and the subsequent development of the formalism by him and other brilliant physicists in the mid-1920s, it became clear that quantum mechanics is not just an evolution, but a scientific revolution. Its startling implications are not only of fundamental importance for most of physics, but also for our understanding of Nature and the philosophy of science in general. In addition, they paved the way for the theoretical and technological achievements in the newer topics of quantum information, communication, and computer science with its now omnipresent effects on our everyday lives.

In this contribution I'll try to review in a concise style the basic aspects of the {\scshape Hilbert}-space formalism for quantum systems in terms of its structural elements, its main notions, and its most important mathematical consequences. To this end, I start out in the next section – making up roughly one half of the text – with comparing \emph{canonical} classical (statistical) mechanics to the corresponding quantum mechanics, with an emphasis on the latter. This seems to be reasonable, because (i) it reflects the historical development and (ii) most of the mentioned quantum notions and consequences retain their significance for non-mechanical quantum systems such as pure spin systems, which only appear in the subsequent sections.

Two remarks are in order. First, in his development of the {\scshape Hilbert}-space formalism {\scshape von Neumann} did not just rely on the analogy to classical mechanics, but aimed for a quantum formalism which could stand on its own, see \cite{vN2018,DJ2013}. For this reason, he can also be considered one of the founding fathers of the newer topics \cite{Au2007,SS2008,Pet2008,NC2010,Hay2017} mentioned above.\footnote{But not discussed in this text, for lack of space and because I am only faintly familiar with them.} 
Second, for their development it is typically sufficient to deal with a {\scshape Hilbert} space of \emph{finite} dimension, that is, with linear algebra and its associated matrix theory \cite{Hal1987}. In contrast, {\scshape Heisenberg} and the other early giants of quantum mechanics were from the very beginning confronted with unbounded operators acting on an infinite-dimensional {\scshape Hilbert} space with its topological intricacies and allowing for continuous spectra. Only amazing intuition enabled them to overcome these complications on their way to creating a fundamentally new physical theory and inspiring {\scshape Hilbert}, {\scshape Weyl}, {\scshape Stone}, and other great mathematicians to develop \emph{functional analysis} further. In fact, the first rigorous proof of the \emph{spectral theorem} for general self-adjoint operators was given only about 1930 by {\scshape von Neumann}, who was a visionary polymath \cite{Bha2021}. 

The following sections are mainly meant for readers who want to grasp, or remember, the essential concepts and consequences without getting lost in a morass of technicalities\footnote{Although – quoting Kai Lai {\scshape Chung} – one person's technicalities are another person's profession.}, when having to accept a {\scshape Hilbert} space of infinite dimension, like for the spectral theory of the \emph{hydrogen atom} \cite{Pa1926,Sc1926a,GP1990,Boh1993} or for general \emph{scattering theory} \cite{Bo1926,GP1991,Boh1993,Ya2000}. Accordingly, I often do not adhere to the standards of mathematical rigor. In particular, domain questions of unbounded operators are almost always ignored. Readers interested in more mathematical details
may consult \cite{BEH2008,T2014,Mor2017} or \cite{Stro2008,GS2020}. For readers wanting to understand, how the quantum formalism was (hypothetically) deduced from experiments and intuition I recommend \cite{LévB1990}. In the supporting Biblio\-graphy at the end, some of the selected books are not cited in the main text. But every single one of them, cited or not, has something special to earn the attention of the reader. Of course, there are several other books of this kind.

\setlength{\parindent}{0pt}

\section{Canonical Classical Mechanics and Quantum Mechanics in Comparison}

Classical \emph{statistical} mechanics (\CM) and quantum mechanics (\QM) have many structural similarities. Nevertheless, there are fundamental differences due the \emph{non-commutativity} in the quantum structure discovered in 1925 by {\scshape Heisenberg} \cite{Hei1925,Hei1971,BJLR2017}.
Their main consequences have then been worked out jointly with {\scshape Born} and {\scshape Jordan} \cite{BJ1925,BHJ1926}. Other highly important contributions to the quantum formalism are due to {\scshape Dirac} \cite{Di1925}, {\scshape Schrödinger} \cite{Sc1926a,Sc1926b}, and {\scshape von Neumann} in his classic \cite{vN1932,vN2018}.          

\subsection{Structural elements of classical and quantum mechanics – contrasted in tabular form}

It is sufficient and convenient to consider the simple model system ``point mass in the {\scshape Euclid}ean line $\R$'' for a particle with mass $m>0$, but ``practically'' without spatial extent (and spin), moving rectilinearly at (variable) speed much smaller than that of light in vaccum.
\begin{small}
\begin{center}
 \begin{tabular}{p{0.26\textwidth}|p{0.32\textwidth}p{0.31\textwidth}}\hline\\[-0.3em]
&Classical Mechanics (\CM)&Quantum Mechanics (\QM)\\[0.3em]
\hline\\
$\bullet$ Arena (or Stage) &phase space~$\Gamma$\newline [here: $\qquad\Gamma = \R \times \R$\,]&separable {\scshape Hilbert} space $\H$\newline over\;the\;complex\;numbers\;$\C$ [here: $\qquad\H=\LL^2(\R)$\,]\\[3em]
$\bullet$ Canonical Elements \newline (momentum and position)&real coordinates of $\Gamma$ &basic self-adjoint operators on $\H$\newline\\[-0.4em]
&\centerline{$p,q$}&\centerline{$P,Q$}\\[0.7em]
$\bullet$ Elements of the\newline {\scshape Liouville} Space & complex functions $\,a:\Gamma\to\C$ \[a(p,q)\] &operators $\,A$ on $\H$ \[A(P,Q)\]\\[-1.5em]
$\bullet$ Scalar Product &\centerline{$\displaystyle{\langle a, b\rangle  \ceq \int_\Gamma \frac{\dd p \,\dd q}{2\pi\hbar} a^*(p,q)\,b(p,q)}$} &\centerline{$\displaystyle{\langle A,B\rangle  \ceq \tr(A^*B)}$}\\[1.5em]
$\bullet$ {\scshape Lie} Product &{\scshape Poisson} bracket \[ [a,b] \ceq \frac{\partial a}{\partial p} \frac{\partial b}{\partial q} - \frac{\partial b}{\partial p} \frac{\partial a}{\partial q} \] &(standardized) commutator \[ [A,B] \ceq\frac{\mathrm{i}}{\hbar} (AB-BA) \]\\[-0.3em]
$\bullet$ Canonical Relation &\centerline{$\displaystyle{[p,q]=1}$} &\centerline{$\displaystyle{[P,Q] = \uu}$}\\[1em]\hline\\
$\bullet$ Observables & real ({\scshape Borel}) functions & self-adjoint operators \\[0.5em]
&\centerline{$\displaystyle a^*=a $} &\centerline{$\displaystyle A^*=A $}\\[-0.5em]

$\displaystyle\quad\text{\parbox{2.8cm}{\footnotesize Important example:\newline Standard {\scshape Hamilton}ian}}$ &\centerline{$\displaystyle h(p,q) = \frac{1}{2m}p^2 + \V(q)$} &\centerline{$\displaystyle\qquad H = \frac{1}{2m} P^2 + \V(Q)\displaystyle\text{\phantom{\parbox{1cm}{AAA AAA AAA}}}$}\\[-0.3em]
&\multicolumn{2}{c}{\footnotesize ``kinetic plus potential energy''}\\[0.7em]
$\bullet$ Events & indicator functions of\newline ({\scshape Borel}) subsets $\mathcal{E} \subseteq \Gamma$ 
\[ \chi_\mathcal{E}(p,q) \ceq \Big\{\begin{array}{ll}1\, \quad &\text{if}\; (p,q) \in   \mathcal{E}\\   0\, &\text{else}\end{array}\] &projections on (closed)\newline subspaces $E\H\subseteq \H$ \[E^* = E = E^2\]\\\hline
\end{tabular}

\bigskip
\begin{tabular}{p{0.26\textwidth}|p{0.32\textwidth}p{0.31\textwidth}}
\hline\\[0.2em]
   
   $\bullet$ States &probability densities on $\Gamma$\newline (possibly\,in\,the\,distributional\,sense)  &statistical operators on $\H$ \newline (still often: ``density matrices'') \newline\\
&\centerline{$w^* = w \geq 0\,, \quad   \langle w, 1 \rangle  = 1$} & \centerline{$W^* = W \geq 0\,, \quad \langle W,\uu\rangle  = 1$}\\[1em]

   $\bullet$ Pure States &\centerline{{\scshape Dirac} measures \text{at some}\,$\displaystyle (p_0,q_0)\in\Gamma$}\smallskip
\centerline{$\displaystyle w(p,q) = 2\pi \hbar\,\delta(p-p_0)\,\delta(q-q_0)$} &\centerline{$\displaystyle W = W^2$}\\[2.5em]
   $\bullet$ Entropy of a State &\centerline{$- \langle w, \ln w \rangle$} &\centerline{$- \langle W,\ln W \rangle$}\\[1em]\hline\\
   $\bullet$ Eigenstate\newline of an Observable &\centerline {$ w(p,q) = \rho(p,q)\,\delta\big(a(p,q) - \alpha\big)$} \centerline{\;\;\text {with some}\, $\displaystyle\rho \geq 0\,, \quad \alpha \in \R$} &\centerline{$ AW = \alpha W$}\centerline{$\alpha\in\R$}\\[2em]
    $\bullet$ Expectation Value \newline of an Observable in a State &\[\langle  a \rangle  \ceq \langle  w, a \rangle\] &\[\langle A\rangle  \ceq \langle  W,A \rangle\]\\[-0.8em]
   $\bullet$ Probability of an Event\newline Occurring in a State &\[\langle\chi_\mathcal{E} \rangle\]&\[\langle  E \rangle\]\\\hline\\
  
 $\bullet$ Time-Evolution\newline of a State: \medskip\newline instantaneous \medskip\newline prolonged \footnotemark \medskip \newline\ [in the {\scshape Schrödinger} \newline picture]&\centerline{{\scshape Liouville} equation} \[ \frac{\dd}{\dd t} w_t = [w_t, h] \]\[w_t=\text{e}^{-t[h,\boldsymbol{\cdot}]}\,w_0\]
&\centerline{{\scshape von Neumann} equation}\[\frac{\dd}{\dd t} W_t = [W_t, H] \]\[W_t=\text{e}^{-\ii t H/\hbar}\,W_0\, \text{e}^{\ii t H/\hbar}\]\\\hline\\

   $\bullet$ Invariance Group &canonical transformations & (anti-)unitary transformations\\[1em]\hline
 \end{tabular}
\end{center}
\end{small}

\vspace{-1em}
\begin{figure}[h!]
\begin{center}
\includegraphics[scale=0.20]{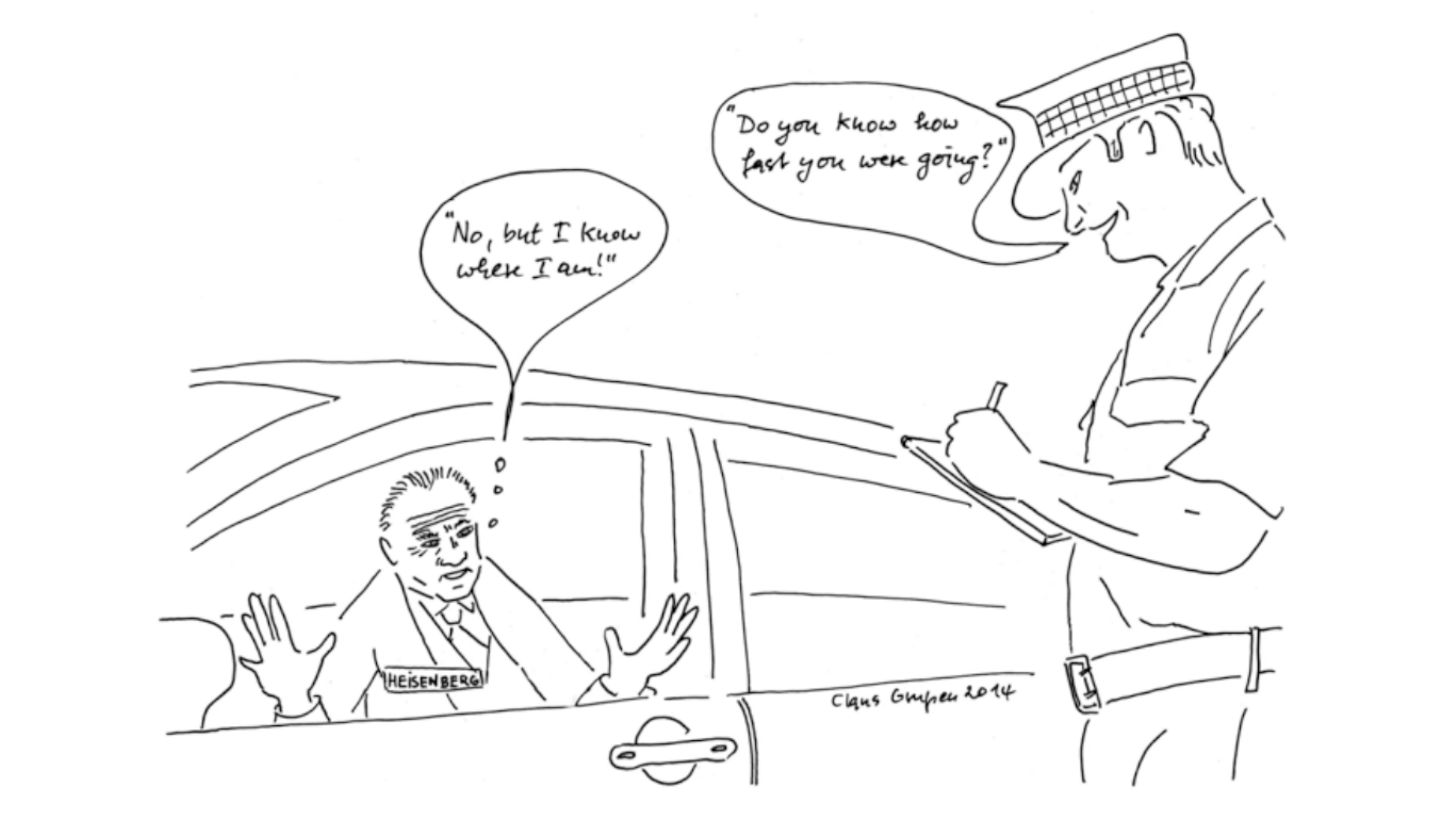}
\end{center}
\vspace{-1.5em}
\caption{\footnotesize \emph{Quantum indeterminacy}\; (cartoon from \cite{Kl2019}, $\copyright$ Claus {\scshape Grupen}\, 2014)}
\end{figure}

\footnotetext{For short times, that is, for small $|t|$, the evolution is (informally) given by the {\scshape Lie} expansion formula:
\[ w_t=w_0-\frac{t}{1!}[h,w_0]+\frac{t^2}{2!}\big[h,[h,w_0]\big]+\dots\quad\text{ resp. } \quad\displaystyle W_t=W_0-\frac{t}{1!}[H,W_0]+\frac{t^2}{2!}\big[H,[H,W_0]\big]+\dots\]
}

\newpage

\subsection{Explanations and comments}

\begin{enumerate}[label=\arabic*)]
	\setlength\itemsep{0.5em}
\item Arena \label{arena}

\begin{itemize}
	\setlength\itemsep{0.25em}
   \item The arena of the \CM\ system is [here] just the {\scshape Euclid}ean plane $\R\times\R\cong\R^2$ with the momentum $p$ (or velocity times mass $m$) and position $q$ serving as (linear orthogonal) coordinates for its points in the sense of {\scshape Cartesius/Descartes}.
    \item The arena of the corresponding \QM\ system is a {\scshape Hilbert} space $\H$ over the (algebraic field of) complex numbers $\C$ of [here] countably \emph {infinite} dimension. This means that $\H$ is a \emph{linear space} of vectors $\psi,\varphi,\eta,\dots$ with a scalar product $\langle\psi|\varphi\rangle$ and norm $\|\psi\|\ceq\langle\psi|\psi\rangle^{1/2}$, where $\langle\psi|\varphi\rangle$ is assumed to be anti-linear in the left argument. Moreover, there is a countable set of pairwise orthogonal unit vectors (orthonormal basis $\eqc$ ONB) such that every $\psi\in\H$ can be written as a linear combination of these unit vectors which converges (in norm). Finally, every {\scshape Cauchy} sequence in $\H$ is convergent. To summarize, $\H$ is the ``natural'' extension to infinite dimension of the $d$-dimensional \emph {unitary} space $\C^d$ with \emph{finite} $d\in\{1,2,3,\dots\}$, the complex analog of the {\scshape Euclid}ean space $\R^d$. Although all {\scshape Hilbert} spaces of the same dimension are isometrically isomorphic, it is convenient throughout this Section 2 to follow {\scshape Schrödinger} \cite{Sc1926a} and to think of $\H=\LL^2(\R)$. This is the linear space of all functions $\psi:\R\rightarrow\C, q\mapsto \psi(q)$ being square-integrable with respect to the {\scshape Lebesgue} measure, that is, $\int_\R \dd q \,\big|\psi(q)\big|^2<\infty$. Here two functions $\psi,\varphi\in\LL^2(\R)$ are identified with each other if their values coincide \emph{almost everywhere} (a.e.), that is, differ at most on a set of {\scshape Lebesgue} measure zero. The scalar product for $\LL^2(\R)$ is given by $\langle\psi|\varphi\rangle=\int_{\R}\!\dd q\,\psi^*(q) \varphi(q)$, where $\psi^*\in\LL^2(\R)$ is defined by pointwise complex conjugation: $\psi^*(q)\ceq\psi(q)^*$. Whether $\LL^2(\R)$ or another realization of $\H$ is used, all elements and notions in the right column of the table that do not refer to $P$ and $Q$ of a \emph {mechanical} system in the strict sense, retain their significance for other quantum systems with infinite-dimensional or finite-dimensional $\H$. For example, for pure spin systems one may think of $\H=\C^d$. 
\end{itemize}

\item Canonical Elements \label{kanonische_variablen}

\begin{itemize}
	\setlength\itemsep{0.25em}
	
  \item For the \CM\ see Explanation~\ref{arena}.
  \item For the \QM\ (in the {\scshape Schrödinger} realization) the \emph{position operator} $Q$ is defined as the basic \emph{multiplication} operator by $(Q\psi)(q)\ceq q\psi(q)$ for all $\psi\in \dom(Q)\ceq\{\eta\in\LL^2(\R) : \|Q\eta\|<\infty\}$, the \emph{domain} of $Q$. It's obvious that $\langle\varphi|Q\psi\rangle=\langle Q\varphi|\psi\rangle$ for all $\varphi, \psi \in \dom(Q)$ (with the finiteness following from the {\scshape Cauchy--Schwarz} inequality $|\langle\varphi|Q\psi\rangle|\leq\|\varphi\|\|Q\psi\|<\infty$). Since $\dom(Q)$ is dense in $\LL^2(\R)$, it is not hard to show that the (unique) adjoint operator $Q^{*}$ has the same domain. Hence $Q$ is \emph{self-adjoint} on $\dom(Q)$ which is written as $Q^{*}=Q$, see \cite{BEH2008,T2014}.\\ The definition of the \emph{momentum operator} $P$ is more involved. Informally, it is the basic \emph{differential} operator ``$-\ii\hbar\frac{\dd}{\dd q}$'', where  $2\pi \hbar\approx 6.63\times10^{-34}$ {\scshape Joule}-seconds is the fundamental {\scshape Planck} constant. Its precise domain turns out to be 
\[\dom(P)\ceq\big\{\eta\in \LL^2(\R): \eta\; \text{is absolutely continuous with derivative}\ \eta'\in \LL^2(\R)\big\}\,.\]
A function $\eta: \R\rightarrow\C, q\mapsto\eta(q)$ is \emph{absolutely continuous} if its (a.e. existing) derivative $\eta'$ satisfies the fundamental theorem of calculus
in the {\scshape Lebesgue} sense
\[\int_{q_1}^{q_2}\dd{q}\,\eta'(q)=\eta(q_2)-\eta(q_1)\quad\text{for all}\;\; q_1, q_2\in\R\;\;\text{with}\;\; q_1\leq q_2\,.\]   
For $\psi\in\dom(P)$ it can be shown that $\psi(q)\to 0$ as $q\to\pm \infty$\,. 
\end{itemize}
\newpage
\item Scalar Product (for the {\scshape Liouville} space) \label{skalarprodukt}

\begin{itemize}

\item
In the \CM\ case, $\langle  a,b \rangle$ is a (positive definite) scalar product in the strict sense only for $a, b\in \LL^2(\R\times\R)$. For other functions on $\R\times\R$  it is merely a convenient notation for the underlying integral if it exists. Simple example: $\langle |a|,1 \rangle <\infty$ and $|b(p,q)|\leq \gamma$ for {\scshape Lebesgue}-almost all $(p,q)\in \R\times\R$ with some constant $\gamma < \infty$.
\item 
Similar remarks apply to the \QM\ case. Here  $\tr(A^*B)$ makes sense as a scalar pro\-duct if $A$ and $B$ are {\scshape Hilbert--Schmidt} operators. But the \emph{trace} also exists, for example, if $A$ is even of the \emph{trace class}, $\tr\sqrt{A^*A}<\infty$, and $B$ is only \emph{bounded} in the sense that its uniform norm $\|B\|\ceq \sup\{\|B\eta\| : \eta\in \H\,, \|\eta\|=1\}$ is finite.
\end{itemize}
  
\item {\scshape Lie} Product \label{lie-produkt}

\begin{itemize}
		\setlength\itemsep{0.25em}
	\item is, by definition, bilinear, anticommutative, and obeys the {\scshape Jacobi}  identity  
	\[ \big[A,[B,C]\big]+\big[B,[C,A]\big]+\big[C,[A,B]\big]=\mathbb{0}\,.\]
	\item is, in addition, compatible with the ``ordinary'' product by the {\scshape Leibniz} rule
\[ [AB,C]=A[B,C]+[A,C]B\,.\]

\end{itemize}
[similar in the \CM\ for differentiable $a$, $b$ and $c$]

\item Canonical Relation\label{cr}

	\begin{itemize}
	\setlength\itemsep{0.25em}
	\item is obvious in the \CM.
	\item is in the \QM\ known as the {\scshape Heisenberg}--{\scshape Born}--{\scshape 	Jordan} \emph{commutation relation}. It follows \emph{informally} by ignoring 			domain questions and using the product rule of calculus. One rigorous 		approach is through the ``unitary version'' of the relation \cite{Wey1927}.
	\end{itemize}

\item Observables \label{observablen}

	\begin{itemize}
	\setlength\itemsep{0.25em}
	\item represent the \emph{quantities} that are \emph{measurable} in the sense of 		experimental physics. Basic examples are the canonical elements themselves. But is 	there an operational instruction to measure, for example, $PQ^2P=QP^2Q$ in the 		laboratory? I don't know.
	\item can be interpreted in the \CM\ directly as real-valued \emph{random variables} in the sense of \emph{classical probability theory} (\CPT) due to {\scshape Kolmogorov} 1933, see \cite{Bau1996}.
	\item manifest themselves in the \QM\ \emph{physically} in that only the (real) \emph{spectral values} of the self-adjoint operator are postulated as possible outcomes of measurements in the laboratory. Thereby they become \emph{quantum} random variables, but in a more fundamental and intrinsic sense than in the \CM. I will try to explain this in the following.
	\end{itemize}

\item Events \label{ereignisse}

	\begin{itemize}
	\setlength\itemsep{0.25em}
	\item are \{1,0\}-(spectral)valued or {\scshape Bernoulli} observables. They correspond to simple \textsf{yes/no} alternatives and form a \emph{lattice} or algebra endowed with a complementation operation, classically even a {\scshape Boole}an one, see Section~4.
	\item are often induced by an observable and a {\scshape Borel} set $I \subseteq \R$ in the following sense:

\smallskip

$\chi_{a^{-1}(I)}$\,=\,indicator function of the pre-image $a^{-1}(I)\ceq\{(p,q)\in \R\times\R: a(p,q)\in I\}$ of $I$\, under $a$\; resp.\;
$\chi_I(A)$\,=\,\emph{spectral projection} of $A$ for $I$\, by the spectral theorem.
\medskip

This is interpreted as follows:

\begin{enumerate}[label=\roman*)]
 \item Mathematically: 
``$a$ resp. $A$ has or has not a (spectral)value in $I$.'' 
 \item Physically:
``{}\,The act of a single (ideal) measurement of $a$ resp. $A$ in the laboratory does or does not give a result in $I$.''{}
\end{enumerate}

\item  may be combined to form a \emph{joint} event, if there are two of them, by the well-known formula  $\chi_{{\cal E} \cap {\cal F}}=\chi_{\cal E}\chi_{\cal F}$ in the \CM\ and \CPT. It generalizes in the \QM\ to $E\wedge F\ceq\lim_{n\to\infty} E(FE)^n$, because in general $EF\neq FE$, see Section 4.
Example:  For the spectral projections $E=\chi_I(P)$ and $F=\chi_J(Q)$ on $\H=\LL^2(\R)$ associated with non-empty and bounded {\scshape Borel} sets $I,J\subset \R$ one finds the \emph{impossible} event $E\wedge F=\mathbb{0}$ (equivalently $E\H \cap F\H =\{0\}$), as a consequence of the ``delocalization property'' of the {\scshape Fourier--Plancherel} transformation, see Sections~2.3 and 4 as well as pp.104--109 in \cite{BGL1995}. In other words, these two events are \emph{disjoint} (or \emph{mutually exclusive}).

\item have a natural generalization to observables taking (spectral) values in the unit interval  ${[0,1]}$. They are mainly of interest in the \QM, where they obey the operator inequali\-ties $\mathbb{0}\leq E \leq \uu$ and form a kind of fuzzy or unsharp events, often called \emph{effects} \cite{Lu1985/87}. More importantly, they are basic for the effect-valued or, less precise,  \emph{positive-operator-valued} (POV) measures on $\R$ and the related \emph{unsharp} observables \cite{BGL1995,BZ2006,Au2007,NC2010,BLPY2016,Hay2017}. These are important for modern measurement theory and quantum information theory, but they are not considered in this text.    
\end{itemize}

\item States \label{zustaende}

\begin{itemize}
	\setlength\itemsep{0.25em}
\item  A \emph{state} steers the randomness of the observables of the system. The notion includes those needed for \emph{statistical} mechanics and for a \emph{composite} system, when restricting/reducing its (total) state to one of its subsystems, see Section 3. A quantum state is a positive operator of trace class and therefore only has a discrete spectrum.

Example: {\scshape Boltzmann}--{\scshape Gibbs} state
 \[ W = \frac{\textrm{e}^{-\beta H}}{\tr \textrm{e}^{-\beta H}}\]
of a quantum system with {\scshape Hamilton}ian $H$ (= energy observable) in thermodynamic equlibrium at absolute temperature  $1/(\beta k_B) > 0$\,. Here the {\scshape Boltzmann} constant $k_B>0$ fixes the {\scshape Kelvin} scale and the discrete spectrum of $H$ is such that the trace in the denominator, the canonical \emph{partition function}, is finite. Example: $H=P^2/(2m)+\gamma\,Q^2$ with $\gamma >0$, defining the quantum \emph{harmonic oscillator}.
\smallskip
\item A \emph{pure} quantum state is a one-dimensional projection:
\[W =|\psi_0\rangle\langle\psi_0| \ceq \text{projection on the span of}\; \psi_0 \in \H\,\,\text{with}\,\, \langle\psi_0|\psi_0\rangle=1\] \[\psi_0 = \text{\emph{state vector}}= \text{ ``{}normalized wave function''{}}\,.\] 
A \emph{pure} classical state is fully determined by only two real numbers $(p_0,q_0)$. In contrast, a pure quantum state needs an infinite sequence of complex numbers for its characterization [given an ONB in $\LL^2(\R)$], many of them possibly equal to $0$.

\item \emph{Mixed} (or \emph{impure}) quantum states are non-trivial (possibly infinite) convex combinations of pure ones according to their \emph{spectral resolution} (by the spectral theorem):
\[ W = \sum_n \rho_n  \,|\psi_n\rangle\langle\psi_n| \]	
\[\rho_n \geq 0,\quad \sum_n \rho_n = 1,\quad \psi_n\in\H,\quad\langle\psi_n| \psi_m\rangle = \delta_{n,m}\,, \quad \sum_n |\psi_n\rangle\langle\psi_n| = \uu\,.\]
\end{itemize}
\newpage
\item Entropy of a State (following {\scshape Boltzmann}, {\scshape Gibbs}, {\scshape von Neumann}, and {\scshape Shannon}) \label{entropie}

\begin{itemize}
	\setlength\itemsep{0.25em}

\item It may be understood purely information-theoretically in the sense of {\scshape C.E. Shannon} and quantifies how much \emph{information} the state lacks compared to the maximum possible, that is, how impure it is. Only for pure states is the entropy minimal. The larger it is, the less (or more coarse-grained) is the information encoded in the state.

\item In the \CM\ there exists, in principle, only subjective ignorance (or ``uncertainty''), whereas in the \QM\ there may be additional objective \emph{indeterminacy}, even for pure states. See Explanation~\ref{varianz}.

\item The entropy of the {\scshape Boltzmann}--{\scshape Gibbs} state of Explanation~\ref{zustaende} equals the \emph{thermal} entropy (of a single particle), when multiplied by $k_B$.

\item Range of the entropy:
\begin{align*}
\text{\CM: } &&- \infty &\leq - \langle w,\ln w\rangle \leq  \ln \Big( \int_\Gamma \frac{\dd p\, \dd q}{2 \pi \hbar} \Big)\quad
\big[=\infty\quad \text{for\,\, $\Gamma=\R\times\R$\big]}\\[0.5em]
\text{\QM: } && 0 &\leq - \ln \langle W,W\rangle\leq -\langle W, \ln W\rangle\\&&&\leq \ln\big(\dim \H \big)\quad\big[= \infty\quad\text{for\,\, $\H= \LL^2(\R)\big]$}
\end{align*}
[The lower bound\footnote{It follows from the spectral resolution of\, $W$ (with the definition\, $0\ln 0\ceq0$) and the {\scshape Jensen} inequality \cite{Bau1996} applied to the (concave) logarithm:\; 
$\tr W\ln W =\sum_n \rho_n\ln \rho_n \leq\ln\sum_n{\rho_n}^2 = \ln\tr W^2\,.$ By the way: Since $\sum_n{\rho_n}^2>0$, the second-order quantum {\scshape R\'enyi} entropy is always finite, even if $\dim \H=\infty$\,.} 
in the \QM\ case is the second-order quantum {\scshape R\'enyi} entropy.]
\end{itemize} 

\item Eigenstate of an Observable

\begin{enumerate}[label = \roman*)]
	\setlength\itemsep{0.25em}
\item  For the \CM\ one has:
\begin{itemize}

\item An eigenstate is concentrated on the level line corresponding to one of the possible values the function $a$ can take. Therefore one has $a\,w = \alpha\,w$ similarly to the \QM.
\item Common eigenstates of two observables only exist if these have level lines crossing each other.
\item Every pure state is a common eigenstate of \emph{all} observables (if defined on the
whole phase space).

\end{itemize}
\item  For the \QM\ one has: 
\begin{itemize}

\item The operator $A$ has eigenstates only if $A$ has \emph{eigenvectors} $\psi\in\H$. Upon normalization they determine the pure eigenstates of $A$ according to: $W=|\psi\rangle\langle\psi|$, $\langle\psi|\psi\rangle=1$, $A\psi=\alpha\psi$\,.

\item Mixed eigenstates of $A$ only exist if the multiplicity of the corresponding eigenvalue is at least $2$; namely as an arbitrary convex combination of pure eigenstates of this eigenvalue.

\item Common eigenstates of two observables exist in general only if these \emph{commute}.

\end{itemize}
\end{enumerate}

\item Expectation Value (mean or average value) of an Observable in a State\label{erwartungswert} 

\begin{itemize} 
	\setlength\itemsep{0.25em}

\item is a linear, real, positive, and normalized functional on the star-algebra of (bounded) operators [resp. functions in the \CM\ and \CPT] in the sense that

\[\langle\alpha A + \beta B\rangle=\alpha\langle A\rangle + \beta\langle B\rangle\quad(\alpha,\beta\in\C)\,,\]
\[\langle A^*\rangle=\langle A\rangle^*\,,\quad\langle A^* A\rangle\geq 0\,,\quad\langle \uu\rangle=1\,.\]

These properties are sufficient \cite{Far1978} to imply  the {\scshape Cauchy}--{\scshape Schwarz} inequality $|\langle A^*B\rangle|^2 \leq \langle A^*A\rangle\langle B^*B\rangle$. 
[similar in the \CM]
\item is in the \CM\ case, $ \langle  a \rangle = \langle w,a\rangle$, independent of the {\scshape Planck} constant $2\pi \hbar$ by the normalization $\langle w, 1 \rangle  = 1$ (unless $a$ depends on it). 

\item is a ``weighted mean'' of the (spectral) values an observable can take (in repeated
measurements). The probability that a particular outcome will occur equals the expectation value of the associated event, see the next item as well as the Explanations~\ref{observablen}, \ref{ereignisse}, and \ref{wsk}.

\item may be written in two simple alternative forms
\[\boxedeqn{\braket{A} =\langle W,A\rangle= \tr(WA) =\sum_n \rho_n\,\langle\psi_n| A\, \psi_n\rangle  =   \sum_j \alpha_j \tr(WE_j)\,,}\]
in case the \QM\ observable $A$, just as $W$, has a purely \emph{discrete} spectral resolution
according to $A = \sum_j \alpha_j E_j $ with $\alpha_j \in \R$, $\alpha_i \neq \alpha_j$ if $i \neq j$, $E_i^* E_j = \delta_{ij} E_j$, and $\sum_j E_j = \uu$, where the trace $\tr E_j$ is the dimension of the $j$-th eigenspace, that is, the multiplicity of the eigenvalue $\alpha_j$. Each of the two forms has an obvious probabilistic interpretation.
In case the spectrum of $A$, as a subset of the real line $\R$, has also or even only, a \emph{continuous} part [such as the operators $P$ and $Q$], then the last sum must be replaced by the integral $\int_\R\mu_A(\textrm{d}\alpha)\,\alpha$\,, see the next Explanation. 
\end{itemize}

\item  Probability of an Event Occurring in a State \label{wsk}

\begin{itemize}
\setlength\itemsep{0.25em}

\item 
Each quantum state $W$ induces a \emph{probability measure} $\mu$ on the lattice/algebra of events $E$ defined by $\mu(E)\ceq\braket{E}=\tr\big(WE\big)$, see Explanation~\ref{ereignisse} and Section 4. The corresponding definition in the \CM\ leads to a probability space in the sense of \CPT. This is not true for the \QM, because the corresponding lattice is not of {\scshape Boole}an type which, in turn, is the reason for the quantum particularities such as interference, violations of {\scshape Bell}-type inequalities, and the like, see \cite{Ac2010}, Sections~5 and 6.

\item 
A given state $W$ steers the randomness of \emph{all} observables through their spectral projections considered as events. More precisely, the probability measure $\mu_A$ on the {\scshape Borel} sets $I$ of the real line $\R$, defined by
\[\mu_A(I)\ceq\mu\big(\chi_I(A)\big)=\braket{\chi_I(A)}\,,\quad I\subseteq\R\,,\]
is the probability \emph{distribution} of the observable $A$ in the state $W$. In close analogy to the \CM\ and \CPT, it contains the complete probabilistic information about $A$.  
{\scshape Born} and {\scshape von Neumann} have postulated $\mu_A(I)$ as the probability that, in the state $W$, the outcome of a measurement of $A$ lies in $I$, confer Explanations \ref{observablen}, \ref{ereignisse}, and Section 4. Informally and by not allowing for a \emph{singularly} continuous spectrum \cite{T2014} of $A$, one can associate to $\mu_A$ the (possibly distributional) probability \emph{density} $\rho_A$ given by \[\rho_A(\alpha)\ceq \frac{\dd}{\dd\alpha} \mu_A\big( ]-\infty,\alpha]\big)=\frac{\dd}{\dd\alpha}\big\langle\chi_{]-\infty,\alpha]}(A)\big\rangle=\langle \delta(\alpha \uu-A)\rangle\geq 0\,,\quad \alpha\in\R\]  
[with respect to the {\scshape Lebesgue} measure] in terms of the {\scshape Dirac} delta. 
Then one may write
\[\langle f(A) \rangle=\int_\R\mu_A(\textrm{d}\alpha)\,f(\alpha)=\int_\R\textrm{d}\alpha\,\rho_A(\alpha)f(\alpha)\;\]
for any $\mu_A$-integrable function $f:\R\to\R$\,,\,
symbolically: \;$``\mu_A(\textrm{d}\alpha)=\textrm{d}\alpha\,\rho_A(\alpha)"$.
\newpage
Example:
$\braket{\chi_I(Q)}=\tr\big(W\chi_I(Q)\big)=\int_I \dd q \,\langle q| W |q\rangle$ is the probability of finding the particle in $I\subseteq\R$. Accordingly, its density $\rho_Q(q)=\langle q| W |q\rangle$ is the ``diagonal'' of the \emph{integral kernel} $\langle q|W|q'\rangle$ or ``position representation'' of $W$ in the sense that $(W\eta)(q)=\int_\R\dd q' \langle q| W |q'\rangle \eta(q'),\, \eta\in\LL^2(\R)$. In particular, for a pure $W=|\psi\rangle\langle\psi|$ with $\psi\in \LL^2(\R)$ and $\langle\psi|\psi\rangle = 1$ one obtains the ``usual'' absolute square of\\ {\scshape Schrödinger}'s wave function:\,  $\rho_Q(q) = |\psi(q)|^2$.
\smallskip
\item 
All probabilities are interpretated as \emph{predictions of relative frequencies} in often repeated measurements at a single (always equally prepared) system or in a simultaneous measurement of a large number of dynamically independent equal copies of the system (= \emph{statistical ensemble}). However, according to an important theorem of {\scshape Ozawa}, an observable that admits a repeatable measurement is a discrete observable \cite{Oz1984}, see also \cite{Ho2001,BLPY2016}.
\end{itemize}

\item \emph{Variance} \label{varianz} of an observable in a state

is the mean-square \emph{deviation} from, or the quadratic \emph{fluctuation} around, the mean of an observable (in repeated measurements). It is most commonly used to quantify the ``width'' of the (distributional) density $\rho_A$ of its probability distribution $\mu_A$:

\[\boxedeqn{\sigma_A^2 \ceq \big\langle\big( A- \langle A\rangle\uu\big)^2\big\rangle} = \int_\R\mu_A(\textrm{d}\alpha)\,\big(\alpha-\langle A\rangle\big)^2= \langle A^2\rangle -\langle A\rangle^2 \,.\]
\paragraph{Facts:}
\parbox[t]{0.85\textwidth}{\begin{enumerate}[label = \roman*)]
	\setlength\itemsep{0.5em}
\item $\sigma_{\lambda A}^2=\lambda^2\sigma_A^2\,,\quad\sigma^2_{A+\lambda\uu}=\sigma^2_A\,;\quad \lambda\in\R\,.$
\item \boxedtext{$\sigma_A^2 \geq 0$} \label{var1}    
\item \boxedtext{$\sigma_A^2 = 0 \; \Longleftrightarrow\;\mu_A\;$ is a {\scshape Dirac} measure $\;\Longleftrightarrow\; W$ is an eigenstate of $A$}\label{var2}
\end{enumerate}}

\medskip
[similar in the \CM\ and \CPT;\, \CM\ observables as random variables]

\begin{proof}
\begin{enumerate}[label = \roman*)]
	\setlength\itemsep{0.5em}
\item obvious by definition.
\item for example, by the operator inequality $\big( A - \langle A\rangle\uu \big)^2\geq \mathbb{0} $\,.
\item The first equivalence is standard by the {\scshape Chebyshev} inequality \cite{Bau1996}.\\ The equivalence\newline ``$\sigma_A^2=0\;\Longleftrightarrow\; W$ is an eigenstate of $A$''\; can be shown as follows:
\medskip
\begin{enumerate}
\item[``{}$\Leftarrow$''{}:] The assumption $AW = \alpha W$ with  $\alpha \in \R$ implies $A^2\,W = \alpha^2\,W$\;. The claim follows by taking traces.
\item[``{}$\Rightarrow$''{}:] With\; $C\ceq\big(A - \langle A\rangle\uu\big)W^{1/2}\quad$ one has $\quad C^*C = W^{1/2} \big(A - \langle A\rangle\uu \big)^2 W^{1/2}$ 
\newline and therefore $\tr C^*C=\sigma_A^2$. The assumption\, $\sigma_A^2 = 0$\, hence implies

$C=\mathbb{0} \quad \Rightarrow \quad \big(A- \langle A\rangle\uu \big) W = \mathbb{0} \quad \Rightarrow \quad AW = \alpha W\quad\text{with}\quad\alpha \ceq \langle A\rangle.$\qed
\end{enumerate}
\end{enumerate}

\end{proof}

Consequences of iii) with additions:
\smallskip
\begin{itemize}
	\setlength\itemsep{0.25em}

\item Only if the system is an eigenstate of an observable, with eigenvalue $\alpha$ say, then its (renewed) measurement will (again) give $\alpha$ with probability $1$ (\emph{non-randomness} or \emph{absence of fluctuations} in the sense of $\sigma_a^2= 0$ resp. $\sigma_A^2= 0$).

\item For a \QM\ system with $\dim \H \geq 2$ there exists no state, not even a pure one, that is a common eigenstate of \emph{all} observables.\footnote{Since for $\dim \H = \infty$\, there exist also observables $A$ with continuous spectra, this statement reads more generally: There exists no \emph{sequence} of states $(W_n)$ such that $\lim_{n \rightarrow \infty}  \sigma_{A}^2( W_n) = 0$ for \emph{all} $A$. Famous example: For $P$ \emph{and} $Q$ no such sequence exists by the indeterminacy inequality, see Explanation \ref{uu}.} Simple example: If $\psi, \varphi\in\H$ is an orthonormal pair, then $W\ceq|\psi\rangle\langle\psi|$ is an eigenstate of $A\ceq|\varphi\rangle\langle\varphi|\,,$ but not of $B\ceq |\eta\rangle\langle\eta|$ defined by the ``superposition'' $\eta\ceq (\varphi+\psi)/\sqrt{2}$\,, since $\sigma_B^2\equiv\sigma_B^2(W)=1/4$
[Non-existence of \emph{universally fluctuation-free} or \emph{dispersion-less} states].

\item For an individual \QM\ system (with $\dim \H \geq 2$) it is at most acceptable in eigenstates of $A$ to imagine that the system ``possesses'' an eigenvalue of $A$ as an intrinsic \emph{property} pre-existing before a measurement. In all other states the ``value of $A$'' is \emph{objectively indeterminate}, because the outcomes of repeated measurements on the same, always equally prepared, \QM\ system objectively fluctuate with (a statistics obeying)  $\sigma_A^2> 0$\,.

\item If the \QM\ system is in a general state $W$, then the outcome value of an ideal  measurement of a discrete\footnote{for a continuous observable see \cite{Oz1984}.}observable $A$ only \emph{arises} in the moment of measurement, namely \emph{by chance} with the probability $\langle W, E\rangle$. Here $E$ is the eigenprojection of $A$ corresponding to the outcome. According to an ``update rule" of {\scshape Lüders} \cite{Lü1951}, $W$ then changes to the state $W_E\ceq EWE/\langle W, E\rangle$. Such a non-linear and non-invertible change is called a \emph{state collapse}. It is usually undisputed that it occurs in experiments, at least approximately and, in fact, regardless of the knowledge of the persons involved. However, whether it can be explained in a purely quantum-theoretical way is still controversial today, despite the ideas of {\scshape Zeh} \cite{Zeh1970} and subsequent authors to understand it as a consequence of the \emph{entanglement}\footnote{see Section~3.} with the noisy environment (from the measuring apparatus etc.) and the resulting irreversible ``loss of its phase coherence'' in the macroscopic limit (In short: decoherence). See also Section~5.
Apart from this, the state  $W_E$ is (for $\dim\H\geq 3$) the only (!) one that provides for the occurrence of any event $F$ with $EF=F$ ($\Leftrightarrow F\H\subseteq E\H$) in a subsequent measurement the (conditional) probability $\langle W_E,F\rangle$, in particular\footnote{confer the above \emph{absence of fluctuations}.} $\langle W_E, E\rangle =1$. This claim is shown in \cite{BC1981,Hug1989,V2007} to be a consequence of {\scshape Gleason}'s theorem\footnote{see Section~4.}. 
\end{itemize}

\item \emph{Covariance} of two observables in a state\label{kovarianz}\\
\smallskip
quantifies the correlation of their (symmetrized) fluctuations

\begin{eqnarray*}
&\boxedeqn{\tau_{A,B} \ceq \frac{1}{2} \Big\langle\big(A-\langle A\rangle\uu\big)\big( B-\langle B\rangle\uu \big)+\big(B-\langle B\rangle\uu\big)\big(A-\langle A\rangle\uu\big)\Big\rangle} \\&\displaystyle= \frac{1}{2} \langle AB + BA\rangle - \langle A\rangle \langle B\rangle\,.
\end{eqnarray*}  
It should be noted that the definition of the covariance neither needs nor implies the existence of a joint probability distribution of $A$ and $B$. Regardless of this, correlations play a major role in all of science and technology. But in quantum physics they even have a fundamental meaning. See, for example, Section 6.

\paragraph{Facts:}
\parbox[t]{0.85\textwidth}{
\begin{enumerate}[label=\roman*)]
	\setlength\itemsep{0.5em}
\item $\displaystyle \tau_{A,A}=\sigma_A^2\,,\quad\, \tau_{A,B}=\tau_{B,A}\,,\quad\, \tau_{\lambda A,B}=\lambda\, \tau_{A,B}\,,\quad\, \tau_{A+\lambda\uu,B}=\tau_{A,B}\;;\quad \lambda\in\R$.
\label{var}
\item \boxedtext{$\displaystyle\sigma_A^2 + \sigma_B^2 +2 \tau_{A,B} = \sigma_{A+B}^2$} \label{add}
\end{enumerate}}
\paragraph{\phantom{Facts:}}
\parbox[t]{0.85\textwidth}{
\begin{enumerate}[label=\roman*)]
\setcounter{enumii}{2}
	\setlength\itemsep{0.5em}
\item \boxedtext{$\displaystyle\sigma_A^2 \sigma_B^2-\tau_{A, B}^2 \geq \frac{\hbar^2}{4} \braket{[A,B]}^2$} \,\,\emph{Indeterminacy inequality} (\ini) 
\label{ur}
\end{enumerate}}

\medskip
[similar in the \CM\ and \CPT\,, but with $0$ instead of the commutator term]

\begin{proof}
 \ref{var} obvious by definition.
 For \ref{add} and \ref{ur} consider at first the case $\langle A\rangle = 0$ and $\langle B \rangle = 0$. The expectation value of $A^2 + B^2+AB+BA = (A+B)^2$ then gives \ref{add}. The {\scshape Cauchy}--{\scshape Schwarz} inequality from Explanation \ref{erwartungswert} yields \ref{ur} according to
       \begin{align*}
        \langle A^2\rangle \langle B^2\rangle &\geq \big| \langle AB\rangle\big|^2 
	= \big(\re \langle AB\rangle \big)^2 + \big( \im \langle AB\rangle\big)^2\\
        &= \Big\langle\frac{1}{2} \big(AB+BA\big)\Big\rangle^2 + \Big\langle\frac{1}{2 \ii} 
        \big(AB-BA\big)\Big\rangle^2 =\; \tau_{A,B}^2 + \frac{\hbar^2}{4} \big\langle[A,B]\big\rangle^2.
       \end{align*}
If $\langle A\rangle \neq 0$ and/or $\langle B\rangle \neq 0$, then replace $A$ with $A- \langle A\rangle\uu$ and/or $B$ with $B- \langle B\rangle\uu$, which gives the then correct variances and covariance, but does not change the commutator.
\qed
\end{proof}

\item Explanation and comments on the indeterminacy inequality (\ini)\label{uu}
 
\begin{itemize}
	\setlength\itemsep{0.25em}
\item The \ini\ is often still listed under the imprecise or incorrect designations ``uncertainty relation'' or ``uncertainty principle''.
\item The \ini\ refers to the statistics of outcomes of measurements and is, in this general form (but initially only for pure states), due to {\scshape Robertson} \cite{Rob1929,Rob1930,Rob1934} and {\scshape Schrödinger} \cite{Sc1930,Sc2008}. The \ini\ has nothing to do with possible perturbations of the \QM\ system or with imperfections of single measurements. Instead, it tells the physicists that the {\scshape Hilbert}-space formalism of the \QM\ does not allow for states violating it. This is in accord with {\scshape Einstein}'s thought-provoking ``Erst die Theorie entscheidet darüber, was man beobachten kann.''\footnote{In a conversation with {\scshape Heisenberg} in April 1926, quoted after the latter's autobiography \cite{Hei1969}. In English: ``It is the theory which decides what we can observe'', see p. 63 in \cite{Hei1971} and also \cite{Hol2000,Kl2019}. One should note that {\scshape Einstein} used the term ``theory'' in the hypothetico-deductive sense \cite{Ei1919}. Thereby he anticipated {\scshape K. Popper}'s main idea on the progress of natural science (In short: ``trial and error'').}

\item If $\big\langle[A,B]\big\rangle=0$, then the \ini\ simplifies to the ``classical''
positivity of the ``reduced variance product'' (= determinant of the covariance matrix), $\delta^2\ceq\sigma_A^2 \sigma_B^2-\tau_{A, B}^2\geq0$, well-known in the \CPT\ and \CM\ . If $\big\langle[A,B]\big\rangle\neq0$, then $\delta^2>0$ and, hence, $\sigma_A^2 \sigma_B^2 >0$. Both strict positivities indicate interdependent \emph{quantum} indeterminacy. The smaller $\delta^2$ can be, the higher the possible accuracy, or ``sharpness'', with which the physical values of $A$ and $B$ are \emph{jointly} determined in the given state.
\item The \ini\ is often weakened by omitting $\tau_{A,B}^2$\,, although it may be strictly positive itself. Thereby one (carelessly) also loses the classical inequality.
\item Its best known example is the ``canonical'' one. It goes back to {\scshape Heisenberg} \cite{Hei1927,WF2019} as an approximate measurement-uncertainty equality for {\scshape Gaussian} pure states and to {\scshape Kennard} \cite{Ke1927}, barely four months later,   as a precise inequality for the variance product in case of general pure states:
\[\sigma_P^2 \sigma_Q^2 \geq \tau_{P,Q}^2 + \frac{\hbar^2}{4} \geq \frac{\hbar^2}{4} \quad (\text{\emph{independent} of the state}\,\, W).\]
In particular, there is \emph{no} state with $\sigma_P^2 \sigma_Q^2 < \hbar^2/4$, although the momentum variance $\sigma_P^2>0$ and the position variance $\sigma_Q^2>0$  can be arbitrarily small \emph{individually} (for a state with a small enough width of its probability density $\rho_P$ resp. $\rho_Q$ on momentum resp. position space). For rigorous {\scshape Heisenberg}-type error-disturbance relations see \cite{BLW2014} and \cite{Oz2003,Oz2019}. The authors do not seem to agree on all aspects, but \cite{Er2012} reports experimental spin measurements in support of the latter author.  

\item In the \ini\ for $P$ and $Q$ \emph{equality} holds, for example, for the pure state 
$|\psi_0\rangle\langle\psi_0|$ defined by the complex-valued {\scshape Gaussian} function
\[\psi_0(q)\ceq (2\pi \alpha^2)^{-1/4}\textrm{e}^{-q^2/(4\alpha^2)}\textrm{e}^{\ii\gamma q^2/\hbar}\quad\quad\quad (q,\gamma \in\R,\;\alpha^2>0\,)\,.\]

This state has the covariance  $\tau_{P,Q}=2\gamma \alpha^2=2\gamma\sigma_Q^2$ and a positive {\scshape Wigner} density, see Remark iv) in Section 2.5\,.

\item The analytical power of the ({\scshape Heisenberg}--){\scshape Kennard} inequality must not be overestimated. Lower bounds on the ground-state energy of a standard {\scshape Hamilton}ian are in general better derived by inequalities of {\scshape S.L. Sobolev} or related ones \cite{Far1978}. For example, while it is true that the stability of the hydrogen atom is due to quantum fluctuations, it does \emph{not} follow from $\sigma_P^2 \sigma_Q^2\geq\hbar^2/4$, when generalized to three spatial dimensions.
\end{itemize}

\item \emph{Correlation coefficient} \label{korr} (as ``de-dimensionalized'' covariance)

\[\boxedeqn{ \kappa_{A,B} \ceq \frac{\tau_{A,B}}{\sigma_A \sigma_B}}\qquad (\text{if}\quad\sigma_A \sigma_B > 0)\,.\]

with the ``standard deviations'': $\quad \sigma_A \ceq \sqrt{\sigma_A^2} \geq 0$ \quad and \quad $\sigma_B\ceq\sqrt{\sigma_B^2}\geq 0$\,.

\paragraph{Facts:}
\parbox[t]{0.85\textwidth}{
\begin{enumerate}[label = \roman*), ref=\roman*)]
		\setlength\itemsep{0.5em}
\item\label{kor_def_i} Symmetry: $\quad\kappa_{A,B} = \kappa_{B,A}$

\item\label{kor_def_ii} Invariance under scaling and translation:

$A\mapsto \lambda A\,,\quad\lambda>0\,; \quad A\mapsto A+\xi\uu,\quad\xi\in\R$\,;\quad
similar for $B$

\item\label{kor_def_iii} Indeterminacy inequality: $\quad\displaystyle\kappa_{A,B}^2 \leq 1- \frac{\hbar^2}{4}\big\langle[A/\sigma_A,B/\sigma_B]\big\rangle^2 \leq 1$

\item \label{kor_def_iv} States for the two extreme values:

\boxedtext{$\kappa_{A,B} = \pm 1 \quad \Longleftrightarrow \quad$\parbox[t]{0.55\textwidth}{There exists $\lambda > 0$ such that $W$ is an eigenstate of $A \mp \lambda B$}}\\

Interpretation: 

In the state $W$ the observables $A$ and $B$ are \emph{perfectly} $\begin{cases}\text{correlated.}\\\text{anticorrelated.}\end{cases}$ 

Moreover, in this state one obviously has by iii): \; $\big\langle[A,B]\big\rangle = 0\,.$
\end{enumerate} }

\medskip

\begin{proof} 
\ref{kor_def_i} and \ref{kor_def_ii} are obvious by definition.\\ 
\ref{kor_def_iii} is only a reformulation of the \ini\ from Explanation \ref{kovarianz}. \\ 
It remains to show the equivalence \ref{kor_def_iv}. This can be done ``quite classically'' as follows:

 ``$\Leftarrow$'':\, 
\begin{minipage}[t]{0.885\textwidth} Fact~\ref{var2} for the variance, Facts ~\ref{kor_def_i} and \ref{kor_def_ii} for the covariance, and the inequality of the arithmetic and geometric means yield
\[0 = \sigma_{A \mp \lambda B}^2 = \sigma_A^2 + \lambda^2 \sigma_B^2 \mp 2 \lambda \tau_{A,B} \geq 2 \sqrt{\sigma_A^2 \lambda^2 \sigma_B^2} \mp 2 \lambda \tau_{A,B} = 2\lambda \left( \sigma_A \sigma_B  \mp \tau_{A,B} \right),\]
hence $\sigma_A \sigma_B \leq \pm \tau_{A,B}$. On the other hand, iii) obviously implies $|\tau_{A,B}|\leq \sigma_A \sigma_B\,.$ Thus $\tau_{A,B} = \pm  \sigma_A \sigma_B,\, \text{hence}\,\, \kappa_{A,B} = \pm  1\,.$
\end{minipage}

\medskip
``$\Rightarrow$'':\,
\begin{minipage}[t]{0.885\textwidth} With $\displaystyle \lambda \ceq \sigma_A/\sigma_B >0$ one has
   \[ 0 = \left(\sigma_A - \lambda \sigma_B\right)^2 = \sigma_A^2 + \lambda^2  \sigma_B^2 - 2 \lambda \sigma_A \sigma_B = \sigma_{A \mp \lambda B}^2\;.\]
Here the last equality is a consequence of $\sigma_A \sigma_B =\pm \tau_{A,B}$ (due to the assumption $\kappa_{A,B} = \pm 1$) in combination with the Facts~\ref{kor_def_i} and \ref{kor_def_ii} for the covariance. Fact~\ref{var2} for the variance now shows that the underlying state $W$ is an eigenstate of $A \mp \lambda B$ (with eigenvalue $\braket{A \mp \lambda B}$).
\qed
\end{minipage}
\end{proof}

\item Time-Evolution [of a State] \label{zeitentwicklung}
\smallskip
\begin{itemize}
	\setlength\itemsep{0.25em}

\item is purely deterministic (for a \emph{closed} or isolated system as considered here) and is generated by the time-independent energy observable called {\scshape Hamilton}ian (= {\scshape Hamilton} function $h$ resp. {\scshape Hamilton} operator $H$) according to a canonical resp. unitary transformation. It enables the physicists, in principle, to predict a future \emph{expectation value} for times $t>0$, given the present one at $t=0$, for example, of the position of a particle in \emph{motion}.

\item does not change the entropy, in particular pure resp. mixed states remain pure resp. mixed.

\item is equivalent to the well-known differential equations of {\scshape Hamilton} resp. {\scshape Schrö\-ding\-er} (for pure states).

\item can be ``transferred'' from the state to the observables according to the transformation
\[a_t\ceq\text{e}^{t[h,\boldsymbol{\cdot}]}\,a \quad\quad\text{resp.}\quad\quad A_t\ceq\text{e}^{\ii tH/\hbar}\,A\,\text{e}^{-\ii tH/\hbar}\,.\]
The time-dependence of the physically more important expectation values therefore takes two alternative forms
\[\langle a\rangle_t\ceq\langle w_t, a\rangle=\langle w_0, a_t\rangle\quad\quad\text{resp.}\quad\quad\langle A\rangle_t\ceq\langle W_t, A\rangle=\langle W_0, A_t\rangle\,, \]
known as the ``transition between the {\scshape Schrödinger} and the {\scshape Heisenberg} picture''. However, only in the latter picture one can define and study multi-time correlation functions such as $\tau_{a_{t}, b_{t'}}$ resp. $\tau_{A_{t}, B_{t'}}$ for $t\not=t'$\;.
\medskip
\item is for a \emph{free} particle, classical or quantum, particularly simple in the \newline {\scshape Heisenberg} picture, because \;$H=P^2/(2m)$\; generates the \emph{equations of motion}
\[ P_t=P\quad\quad\text{and}\quad\quad Q_t=Q+\frac{t}{m}P\,.\]
For example, they give for the time-dependence of the position variance and the momentum-position covariance, for an arbitrary initial state $W_0$ with $\langle P^2\rangle <\infty$ and $\langle Q^2\rangle <\infty$, the equations:
\[\sigma^2_{Q_t}-\sigma^2_Q=\frac{2t}{m}\tau_{P,Q}+\frac{t^2}{m^2}\sigma^2_P\quad\quad\text{and}\quad\quad \tau_{P_t,Q_t}= \tau_{P,Q}+ \frac{t}{m}\sigma^2_P\,.\]

They are contained already in \cite{Sc1930,Sc2008}, although {\scshape Schrödinger} did not write down the second one, but see also \cite{Lév1986}. In the case $\tau_{P,Q}<0$, that is, for anticorrelated  $P$ and $Q$ in the initial state, as in the example of Explanation~\ref{uu} for $\gamma<0$, their right-hand sides remain negative only for short enough (positive) times; more precisely, for all $t\in {]0, 2t_0[}$\; resp.\; $t\in [{0, t_0}[$ with $t_0\ceq m|\tau_{P,Q}|/ \sigma^2_P\leq m\sigma_Q/ \sigma_P$. In any case, the ``reduced
variance product'' $\sigma^2_{P_t} \sigma^2_{Q_t}-\tau^2_{P_t, Q_t}$ (see Explanation~\ref{uu}) is independent of $t$. These time-(in)dependencies also hold classically for any initial state $w_0$ with $ \sigma_p \sigma_q \in ]{0,\infty}[ $. They can illustrate the quantum ones by a swarm of dynamically independent classical particles.

\end{itemize}
\medskip
\emph{Time-independent} or \emph{stationary} states (in the {\scshape Schrödinger} picture)
\begin{itemize}
\item are characterized by a vanishing {\scshape Lie} product with the {\scshape Hamilton}ian: 
\[[w_0, h] = 0 \quad\quad\text{resp.}\quad\quad  [W_0, H] = \mathbb{0}\,.\]
Example: {\scshape Boltzmann--Gibbs} state from Explanation~\ref{zustaende}.
\item include in the \QM\ all energy eigenstates, but not so in the \CM.
\item are mixed in \CM\ cases of physical interest.
\end{itemize}

\medskip
\item Invariance Group \label{invarianzgruppe}

Leaves expectation values \emph{numerically invariant}, but {\scshape Lie} products and equations of motion (that is, equations for the time-evolution of states, observables, and expectation values) are in general only \emph{form-invariant}. Since the two invariance groups are not isomorphic, there is no unique prescription to ``quantize'' a given classical canonical system. Related problems arise from the multiplicity of ordering the factors in products of non-commuting operators \cite{MS2018} and from finding domains of (essential) self-adjointness for the {\scshape Hamilton}ian and other operators that capture the underlying physical scenario. For some discussions see Ch.8.3 in \cite{BEH2008}.

\end{enumerate}

\subsection{Entropic indeterminacy inequality for momentum and position}

This subsection may be viewed as a kind of addendum to Explanation \ref{uu} in Section 2.2\.. The variance of a probability density provides just one way to quantify its width.
Another quantifier is its entropy. So it might not come as a surprise that the famous
{\scshape Heisenberg--}{\scshape Kennard} indeterminacy inequality $\displaystyle\hbar/2\leq\sigma_P \sigma_Q$  has an entropic analog. It was conjectured in 1957 by {\scshape Hirschman}\cite{Hir1957} and proved in 1975 by {\scshape Beckner}\cite{Bec1975} and {\scshape Białynicki-Birula--Mycielski}\cite{BB--M1975}, see also \cite{FoSi1997}.

\paragraph{Assumption and notations:}

\begin{itemize}
  \item  Given a unit vector $\psi$ underlying a pure state (position representation)
\[\psi \in \LL^2(\R)\,,\,\,\langle\psi|\psi\rangle=\int_\R \dd q \,\big|\psi(q)\big|^2 = 1\quad\big(|\psi(q)|^2 = \langle q|W|q\rangle\,,\quad W\ceq\big|{\psi}\rangle\langle\psi\big|\big).\]
   
\item Transformed vector $\widehat{\psi}$\, by {\scshape Fourier--Plancherel} (momentum representation)
    \[\widehat{\psi} (p) \ceq \int_\R \frac{\dd q}{\sqrt{2\pi\hbar}}\,\psi(q)\,\textrm{e}^{-\ii pq/\hbar}, \quad \int_\R \dd p \,\big| \widehat{\psi}(p) \big|^2 = 1\,.\]
    
\item  \emph{Pseudo-classical} state $w$ assigned to $W$
     \[{w(p,q) \ceq 2\pi \hbar \,\big|\widehat{\psi}(p)\big|^2 \,\big|\psi (q) \big|^2}\,\]
     \[w \geq 0\,,\quad\langle w, 1\rangle = \int_{\R\times\R} \frac{\dd p\,\dd q}{2\pi\hbar}\, w(p,q) = 1\,.\]
 \end{itemize}
 
\paragraph{Assertion} (\emph{entropic} \ini\ for momentum and position):

\medskip
The \emph{classical} entropy of $w$, in other words, the \emph{pseudo-classical} entropy of $W$, has the state-independent \emph{lower} bound\,
  $\displaystyle\ln \frac{\textrm{e}}{2} $, that is
\[ \ln \frac{\textrm{e}}{2}\leq - \langle w, \ln w\rangle\,.\]

\paragraph{Interpretation:}

\medskip
Although the entropies of the (suitably de-dimensionalized) quantum probability densities on momentum and position space,\, $\rho_P$ resp. $\rho_Q$\,, may be arbitrarily small \emph{individually} and even negative, their \emph{sum} never falls below\, $\ln (\textrm{e}/2)= 1 - \ln 2\ > 0.3068$\,.

\begin{proof} (following \cite{Hir1957,Bec1975,BB--M1975})

The function $f:{[2,\infty[}\rightarrow\R$, defined by the difference
\[f(\sv)\ceq \left(\frac{\su}{2\pi \hbar} \right)^{1/(2\su)}\big\|\psi\big\|_\su-\left(\frac{\sv}{2 \pi \hbar}\right)^{1/(2\sv)} \big\| \widehat{\psi}\big\|_\sv\quad \quad \big(\;\sv \geq 2\,,\quad \su\ceq \sv/(\sv-1)\in {]1,2]}\;\big)\;\]

in terms of the usual {\scshape Lebesgue} norms $\|\varphi\|_r \ceq (\int_\R \dd x \, |\varphi(x)|^r)^{1/r} $\, with\, $r\in [1,\infty[$\,, obeys $f(2)=0$ by the well-known  {\scshape Parseval--\scshape Plancherel} equality, as used already above for $\langle w, 1\rangle = 1$. Less well known is the sharp {\scshape Hausdorff}--{\scshape Young} inequality $f(\sv)\geq 0$ for all $\sv\geq 2$, see \cite {LL2001}. The combination of both implies $f'(2+0)\geq 0$  for the (right-sided) derivate of $f$ at $2$. This last inequality turns out to be equivalent to the assertion when using the formula

\[r^2\frac{\dd}{\dd{r}}\big\|\varphi\big\|_r=  \big\|\varphi\big\|_r^{1-r}\int_\R \dd x |\varphi(x)|^r\ln\big(|\varphi(x)|^r/ ||\varphi ||_r^{r}\big)\,\]

with $r=2$ for $\varphi=\psi$ as well as for $ \varphi=\widehat{\psi}$ and noting again that
$||\psi||_2=||\widehat{\psi}||_2=1$.
\qed
\end{proof}

\paragraph{Remarks:}

\begin{itemize}
\setlength\itemsep{0.25em}
\item The entropic \ini\ for momentum and position is extended from pure to mixed states $W\,(\geq W^2\,)$ by using the more general definition
 \[\boxedeqn{w(p,q)\ceq2\pi \hbar \langle p|W|p\rangle\langle q|W|q\rangle\geq 0\,,\quad \langle w, 1\rangle =1}\]
for the pseudo-classical state $w$ assigned to $W$. This claim simply follows from the spectral resolution\footnote{see Explanation \ref{zustaende} in Section 2.2\,.} of $W$, the convexity of the function $x \mapsto x \ln x$ (for $x\geq 0$), the {\scshape Jensen} inequality \cite{Bau1996}, and the above result for pure states.
\smallskip 
\item The \emph{entropic} \ini\ implies {\scshape Kennard}'s \emph{variance} \ini\,, see Explanation \ref{uu} in Section 2.2, because the pseudo-classical entropy of $W$ can simply be bounded from \emph{above}:
\[\boxedeqn{\displaystyle\ln \frac{\textrm{e}}{2} \leq - \langle w, \ln w\rangle\leq\ln \left( \frac{\textrm{e}}{\hbar} \sigma_P \sigma_Q \right) \,.}\]

\begin{proof}
It's enough to recall the {\scshape Gibbs}-type inequality\footnote{Obviously, it is due to the elementary inequality $d(x,y)\ceq  x\ln x-x\ln y-	x+y\geq 0$ for $x,y\geq 0$. The latter results from $d(0,y)=y\geq 0$ and
$d(x,y)= x[(y/x)-1-\ln 	(y/x)] \geq 0$ for $x>0$ because\, $z-1\geq\ln z$\, for $z\geq 0$. Historical references for the {\scshape Gibbs} inequality and its quantum version can be found in \cite{Fal1970,Hub1970}.}
$-\big\langle w, \ln w\big\rangle\leq-\big\langle w, \ln \widetilde{w}\big\rangle$
for any probability density (or classical state) $\widetilde{w}$ on the phase space $\R   	\times \R$\,.
The claimed upper bound follows from computing $-\big\langle w,\ln \widetilde{w}\big\rangle$ for the bivariate {\scshape Gauss}ian density 
\[\widetilde{w}(p,q)= \frac{\hbar}{\sigma_P \sigma_Q} \exp\bigg\{-\frac{(p - \braket{P})^2}{2\sigma_P^2}-\frac{(q - \braket{Q})^2}{2\sigma_Q^2}\bigg\}\] 
with zero covariance and the means and variances of $P$ and $Q$ in the given state $W$.
\qed
\end{proof}
\end{itemize}
\paragraph{Conclusion:}~

\smallskip
\boxedtext{\parbox{0.98\textwidth}{If the \emph{entropic} \ini\ is strict, then the pseudo-classical entropy sharpens the famous \emph{variance} \ini. Examples:
(i) $|\psi_0\rangle\langle\psi_0|$ with $\gamma\neq 0\,$, (ii) the harmonic oscillator at non-zero temperature, $\beta<\infty$ [For these states see Explanation \ref{uu} resp.\,\ref{zustaende}]. Conversely, equality in the variance \ini\  implies that in the entropic \ini. Examples: (i)\, $\gamma=0$\,,\, (ii)\, $\beta\rightarrow\infty\,$.}}
\paragraph{Remark:}
The entropic \ini\ is complemented by the inequality $ -\langle W, \ln W\rangle \leq-\big\langle w, \ln w\big\rangle$, also valid for any state $W$. It's easier to prove \cite{FL2012}, but for \emph{mixed} $W$ not less interesting. Weakening it by the last inequality from above gives an upper bound\footnote{A sharper bound,\,
$\ln(\textrm{e}/\gamma)$\;with $\gamma\ceq 2\sinh(\hbar/(2\sigma_P \sigma_Q))$,\,results directly from the \emph{quantum} {\scshape Gibbs} inequality, $-\langle W, \ln W\rangle \leq-\big\langle W, \ln \widetilde W\big\rangle$, by the choice\, $\widetilde W=\,\gamma\, \exp \{-(P-\braket{P}\uu)^2/2\sigma_P^2\, -\,(Q-\braket{Q}\uu)^2/2\sigma_Q^2\}$. The inequality itself, also known as the \emph{positivity of relative entropy}, basically follows from the spectral re\-solutions of $W$ and $\widetilde W$, combined with the elementary inequality in footnote 13. This first proof goes back to \cite{DM1936}, as I have learned from Albrecht {\scshape Huber} (Universität Kiel, Germany); see his review \cite{Hub1970}.} on the entropy $s$ of $W$. Written in the form $\sigma_P \sigma_Q\geq \hbar\, \textrm{e}^{\,s-1}$, it is seen to sharpen the variance \ini\ if $s>\ln(\textrm{e}/2)$. For a lower bound on $s$ in terms of a phase-space integral see Remark iii) in Section 2.5\,.
\vspace{-0.8em}
\subsection{{\scshape Feynman--Kac} formula and classical bounds on quantum partition functions}
\setlength{\parindent}{1.8em}
The non-commutativity of $P$ and $Q$ in a {\scshape Hamilton}ian such as $H=P^2/(2m) + \V(Q)$ is also reflected by the thermal properties of the underlying physical system, which are determined by the {\scshape Boltzmann}--{\scshape Gibbs} state  
$W= \textrm{e}^{-\beta H}/ \tr \textrm{e}^{-\beta H}$, see Explanation\,\ref{zustaende} in Section 2.2\,.
The derivation is complicated by the fact that the functional equation of the exponential
for complex numbers does not extend to (functions of) non-commuting operators. However, there is a useful method of writing $\textrm{e}^{-\beta H}$ as a (classically) probabilistic average of a product of two factors, one depending only on $P$ and the other only on $Q$. It was invented by {\scshape R.P. Feynman} in his 1942 Princeton Ph.D. dissertation and now often goes under the name ``the path-integral approach to quantum physics‘‘ \cite{Roe1994,FH2010}. After having heard about it, {\scshape M. Kac} found in 1948 a rigorous proof for the case of interest here.\footnote{{\scshape Feynman} was then mainly interested in quantum dynamics, that is, in the informally similar, but mathe\-matically less robust and powerful case with the imaginary $\ii\,t/\hbar$ replacing $\beta>0$, but see \cite{GZ2016}.} This result became known, see pp.115--116 in \cite{Ka1985} and \cite{Roe1994}, as the {\scshape Feynman--Kac} formula ({\scshape F--K} formula), here concisely written as an operator identity:
\[\boxedeqn{\textrm{e}^{-\beta H}=\int\rho{(\dd w)}\,\textrm{e}^{-\ii w(\beta)P/\hbar}\,\exp\Big(-\int_{0}^{\beta}\dd{\tau}\,\V\big(Q+w(\tau)\uu\big)\Big)\,.}
\]
The first integration is over all continuous functions $w:[0,\infty[\rightarrow\R, \tau\mapsto w(\tau)$ on the positive half-line obeying $w(0)=0$, with respect to the (dimensionless) {\scshape Gauss}ian probability measure $\rho$ that is uniquely defined by the
\[\textit{mean}\quad\int\rho{(\dd w)}w(\tau)=0\quad\text{and}\;\; \textit{covariance}\quad\int\rho{(\dd w)}w(\tau)w(\tau^{\prime})=(\hbar^2/m)\,\text{min}\{\tau,\tau^{\prime}\}\,.\] 
In other words, $\{w(\tau)\}_{\tau\geq 0}$ is the \emph{stochastic process} introduced by {\scshape N. Wiener} \cite{Mas1990} in 1923 for modeling the [seemingly] random paths followed by a single particle in the phenomenon of free {\scshape Brown}ian motion [here only considered in one spatial dimension and with diffusion constant  $\hbar^2/(2m)$]. The technical paper \cite{BLM2004}
provides rather mild conditions on the potential $\V$ for the validity of the {\scshape F--K} formula.
Its analytical efficiency is well illustrated by deriving the following (pseudo-)classical bounds on the quantum canonical partition function, or the equivalent ones on the quantum \emph{free energy} $-\beta ^{-1}\ln(\tr\textrm{e}^{-\beta H}).$
\setlength{\parindent}{0em}
 \begin{center}
 \boxedtext{
 \begin{minipage}{0.97\textwidth}
\begin{theorem*} ({\scshape Feynman--Hibbs} 1965, {\scshape Symanzik 1965})

\medskip
The partition function of the {\scshape Hamilton}ian $H\ceq\frac{1}{2m}P^{2}+\V(Q)$ satisfies the inequalities
	\[z(\beta,\beta)\leq\tr\textrm{e}^{-\beta H}\leq z(\beta,0)
	\]
with the (pseudo-)classical partition function 
	\[z(\beta,\tau)\ceq\int_{\R\times\R}\frac{\dd{p}\dd{q}}{2\pi\hbar}\textrm{e}^{-\beta\big(p^{2}/(2m)+\V_{\tau}(q)\big)}
	=\frac{1}{\lambda}\int_{\R}\dd{q}\,\textrm{e}^{-\beta \V_{\tau}(q)}\,\;,
	\]
the \emph{thermal} {\scshape de Broglie} wavelength $\lambda\ceq\sqrt{2\pi\beta\hbar^{2}/m}          $, and a potential being a {\scshape Gauss} transform of the given potential according to:
\[
			\V_{\tau}(q)\ceq\int_{\R}\frac{\dd{\xi}}{\sqrt{2\pi}}\textrm{e}^{-\xi^{2}/2}\,\V\Big(q+\xi\sqrt{\tau\hbar^{2}/(12m)}\Big)
			=\exp\Big(\frac{\tau\hbar^{2}}{24m}\frac{\dd^{2}}{\dd{q^{2}}}\Big)\V(q)\,,\quad \tau\in[0,\beta]\,.
			\]
\end{theorem*}
\end{minipage}}
\end{center}
\begin{proof}
The F--K formula in the position representation gives for the trace 
\[\tr\textrm{e}^{-\beta H}=\int_{\R}\dd{q}\,\langle q|\textrm{e}^{-\beta H}|q\rangle=\int_{\R}\dd{q}\int\rho{(\dd w)}\,\delta\big(w(\beta)\big)\,\textrm{e}^{-\int_{0}^{\beta}\dd{\tau}\,\V\big(q+w(\tau)\big)}.
 \]
Applying the {\scshape Jensen} inequality to the (convex) exponential with respect to the uniform average $\beta ^{-1}\int_{0}^{\beta}\dd{\tau}(\cdot)$ and interchanging the $\tau$-integration with the two other ones results in
\[\tr\textrm{e}^{-\beta H}\leq\beta ^{-1}\int_{0}^{\beta}\dd{\tau}\int_{\R}\dd{q}\int\rho{(\dd w)} \delta\big(w(\beta)\big) \textrm{e}^{-\beta\V\big(q+w(\tau)\big)}=z(\beta,0)\,.
\]
The equality follows from interchanging the $q$-integration with the $w$-integration and noting the facts $\lambda\int\rho{(\dd w)}\,\delta\big(w(\beta)\big)= 1$ and $\int_{0}^{\beta}\dd{\tau}=\beta$\,. This completes the proof of the upper bound. For the proof of the lower
bound let me associate to each path $w$ its ``time'' average $\overline{w}\ceq\beta^{-1}\int_{0}^{\beta}\dd{\tau}\,w(\tau)$ and define the function $x\mapsto I(x)$ by inserting an additional {\scshape Dirac} delta into the above trace formula
in order to restrict the path integration to a fixed value of $\overline{w}$\,:
\begin{align*}
I(x)&\ceq \int_{\R}\dd{q}\int\rho{(\dd w)}\,\delta\big(w(\beta)\big)\delta\big(q+\overline{w}-x\big)\,\textrm{e}^{-\int_{0}^{\beta}\dd{\tau}\,\V\big(q+w(\tau)\big)}\\&=\int\rho{(\dd w)}\,\delta\big(w(\beta)\big)\,\textrm{e}^{-\int_{0}^{\beta}\dd{\tau}\,\V\big(x+w(\tau)-\overline{w}\big)}\,.
\end{align*}

Clearly, integrating over $x\in\R$ gives the desired trace. However, to simplify the restricted $w$-integration, let me first apply the {\scshape Jensen} inequality to the exponential, but this time with respect to the average $\lambda\int\rho{(\dd w)}\,\delta\big(w(\beta)\big)\big(\cdot\big)$ (over all paths ``returning'' to the origin at ``time'' $\tau=\beta$). Its normalization has already been recognized above. This gives
\[I(x)\geq \frac{1}{\lambda}\exp\big(-\int_{0}^{\beta}\dd{\tau}\,\int_{\R}\dd{y}\,\V(y)\, J(x-y)\big)\]
with the $\tau$-independent {\scshape Gauss}ian probability density on the real line
\[J(z)\ceq \lambda \int\rho{(\dd w)}\,\delta\big(w(\beta)\big)\delta\big(z+w(\tau)-\overline{w}\big)= \frac{\sqrt{12}}{\lambda}\exp\big(-12\pi\, z^{2}/\lambda^{2}\big)\,, \quad z\in\R \,.
\]
The last equality is due to the fact that\,$w(\beta)$\,and $w(\tau)-\overline{w}$\, are two correlated {\scshape Gauss}ian random variables with vanishing means and easily calculable variances and covariance. The claimed lower bound $z(\beta,\beta)$ now emerges from integrating over $x$ (and renaming $x$ to $q$).
\qed
\end{proof}

\paragraph{Remarks:}
\begin{itemize}
	\setlength\itemsep{0.25em}
\item
The above proof of the upper bound $z(\beta,0)$ is due to \cite{Sy1965} confirming an unpublished conjecture of {\scshape Feynman}. Another proof follows from the more general {\scshape Golden-Thompson} inequality $\tr\textrm{e}^{A+B}\leq \tr(\textrm{e}^{A}\textrm{e}^{B})$ of 1965. For this and other trace inequalities see \cite{Si2005}. In any case, the tacit assumption $z(\beta,0)<\infty$ is seen to be a simple sufficient condition for $H$ to only have a discrete spectrum.
\item
The proof of the lower bound $z(\beta,\beta)$ is a streamlined version of the one given in Ch.11-2 of \cite{FH2010}. The {\scshape Gauss} transform $\V_{\beta}$ of the given potential $\V=\V_{0}$\, in $z(\beta,\beta)$ is an \emph{effective} potential in the sense that it takes into account some quantum effects in the classical arena. The less $\V(q)$ changes with $q$ on the scale given by $\lambda$, the sharper the bound is. In any case, it yields the correct high-temperature asymptotics ($\beta\downarrow 0$).
\item
The classical partition function $z(\beta,\tau)$ with the transformed (and thereby ``smoothed'') potential $\V_{\tau}$ for general $\tau\geq0$\ is decreasing in $\tau$. This follows from the semigroup property of the {\scshape Gauss} transformation and the {\scshape Jensen} inequality, see \cite{LeW1989}. For a sufficiently smooth potential $\V$ one even has strict monotonicity,
\[\frac{\partial}{\partial\tau}z(\beta,\tau)=-\frac{\beta\lambda}{48\pi}\int_{\R}\dd{q}\,\textrm{e}^{-\beta \V_{\tau}(q)}\bigg(\frac{\partial}{\partial q} \V_{\tau}(q)\bigg)^2<0\,,
\]
because $\V$, and hence $\V_{\tau}$, is not constant everywhere by $z(\beta,0)<\infty$. Combining this result with the above bounds and referring to the intermediate-value theorem yields $\tr \textrm{e}^{-\beta H}=z(\beta,\overline{\tau}(\beta))$ with a unique $\overline{\tau}(\beta)\in\; ]0,\beta]$ associated to the given $\beta>0$. Hence, for any temperature and any potential, the quantum partition function equals the classical partition function with a well-defined \emph{temperature-dependent} effective potential. However, $\overline{\tau}(\beta)$ as a functional of $\V$ is not explicitly known for a general $\V$. 
\item The lower bound $z(\beta,\beta)$ can be sharpened by applying the {\scshape Jensen} inequality to fewer paths than in the above proof. For an impressive example along these lines see \cite{GT1985,FK1986}. Reviews on such effective-potential methods are \cite{LeW1989,CGTVV1995}.
\end{itemize}
%
\subsection{{\scshape Weyl}--{\scshape Wigner} mapping}
%
This mapping provides a \emph{linear} and injective (or one-to-one) assignment between operators $A$ acting on the {\scshape Hilbert} space $\LL^2(\R)$ and complex-valued functions $a$ defined on the phase space $\R \times \R$. It thereby establishes a \emph{quantitative correspondence} between the \QM\ and \CM\ in the same arena rather than a mere confrontation of respective structural elements \cite{Wey1927,Wig1932,Moy1949}. As a general reference I recommend \cite{CFZ2019} and for more mathematical aspects \cite{dG2017}. 
\paragraph{Definition:}

The {\scshape Weyl}--{\scshape Wigner} \emph{mapping} $A \mapsto a$ is defined by
 \[
  \boxedeqn{a(p,q) \ceq 2 \int_\R \dd r\, \textrm{e}^{2\ii p r/ \hbar} \bra{q-r} A \ket{q+r}\,,}\quad(p,q) \in \R \times \R\,,
 \]
with $\langle q|A|q'\rangle$ denoting the (possibly distributional) integral kernel or position representation of $A$. 
The phase-space function $a$ is called the {\scshape Weyl}--{\scshape Wigner}   \emph{symbol} or, more briefly, the \emph{symbol} of the operator $A$. It may depend on $\hbar$.

\paragraph{Facts:}
 \begin{enumerate}[label = \roman*)]
	\setlength\itemsep{0.25em}
 \item Linearity:\quad $\alpha A + \beta B \mapsto \alpha a + \beta b\quad$ for all\,
  $\alpha, \beta \in \C$, in particular $ \mathbb{0} \mapsto 0 $
 
 \item Compatibility with conjugation:\quad $A \mapsto a \quad \Longleftrightarrow \quad A^* \mapsto a^*$
  
  \item $f(Q) \mapsto f(q),\; g(P) \mapsto g(p)$ \;for ({\scshape Borel}) functions $\,f,g: \R\to\R$, in particular $\uu\mapsto 1$
  
  \item Isometry:\quad $\braket{a,b} = \braket{A,B}$
  
  \item Product:\quad $\displaystyle AB \mapsto a\star b\ceq a \exp \left(- \frac{\ii \hbar}{2} \mathcal{D} \right) b\qquad$
   with\footnote{The phase-space function $a\star b$ is known as the \emph{star product} of $a$ and $b$. The arrows in $\mathcal{D}$ indicate, whether (in the informal {\scshape Taylor} series of the exponential) the differentiation refers to $a$ or $b$. Example:\, $a\mathcal{D} b=[a,b]$.}$\qquad\displaystyle \mathcal{D} \ceq \frac{\overleftarrow{\partial}}{\partial p}  \frac{\overrightarrow{\partial}}{\partial q} - \frac{\overleftarrow{\partial}}{\partial q} \frac{\overrightarrow{\partial}}{\partial p} $

  \item The (standardized) commutator is mapped to the {\scshape Moyal} \emph{bracket} :
   \[
    [A,B] \mapsto \{a,b\} \ceq \frac{\ii}{\hbar}(a\star b -b\star a)= a \,\frac{2}{\hbar} \sin \left(\frac{\hbar}{2}  \mathcal{D} \right) b
   \]
   
  \item {\scshape Weyl} \emph{quantization} as the inverse mapping $a\mapsto A$:
     \[
   A = \int_{\R\times \R} \frac{\dd p\, \dd q}{2\pi \hbar}\, \textrm{e}^{\ii(p\uu-P)q/\hbar}\, a\left(p, Q+\frac{q}{2}\uu \right) 
= \left. a \left(\frac{\hbar}{\ii}   \frac{\partial}{\partial q}, \frac{\hbar}{\ii}\frac{\partial}{\partial p}   \right) \textrm{e}^{\ii\left(qP + pQ \right)/\hbar} \right|_{p=0,\,q=0}\\[1em]\]
\[\langle q |A|q'\rangle= \int_{\R}\frac{\dd p}{2\pi\hbar}\,\textrm{e}^{\ii p(q-q')/\hbar}\,a\Big(p,\frac{q+q'}{2}\Big)\;,\quad (q,q')\in\R\times\R\,.\]

\end{enumerate}

\paragraph{Remarks:}
\begin{enumerate}[label = \roman*)]
	\setlength\itemsep{0.25em}
  \item \begin{itemize}[leftmargin=4em,labelsep=1em]	\setlength\itemsep{0.25em}
\item[$ A \mapsto a$] Phase-space \emph{representation} of the \QM\ ({\scshape Wigner} 1932, {\scshape Moyal} 1949)
   \item[$\quad a \mapsto A$] {\scshape Hilbert}-space \emph{representation} of the \CM\ or (canonical) \emph{quantization} prescription of classical phase-space functions ({\scshape Weyl} 1927) with totally symmetrized factor ordering of $P$ and $Q$.
  \end{itemize}
 \item Observables are mapped to observables.
 
Example:\\ For a twice continuously differentiable function $f: \R\to\R$ one has the assignment
\[p^2f(q)\; \longleftrightarrow \; \frac{1}{4} \left( P^2 f(Q) + 2 Pf(Q)P + f(Q) P^2\right) = Pf(Q)P - \frac{\hbar^2}{4}f''(Q)\,.\]
  
\item Positive operators are in general \emph{not} mapped to positive functions and vice versa.\smallskip\\
In particular, the {\scshape Wigner} \emph{density} $w$, that is, the symbol of a \QM\ state $W$ may take also \emph{negative} values. It is therefore, in general, not a probability density on the phase space. Rather, it is similar to a bounded electric-charge density with positive total charge. This is because $ |w(p,q)| \leq 2$ for any $W$ and, by the isometry, also \; $0\leq \langle w,w\rangle=\langle W,W\rangle \leq \langle W,\uu\rangle= \langle w,1\rangle = 1$\;and\, $\braket{w,a} = \braket{W,A}$ for expectation values. This implies that the lower bound on the entropy of a (mixed) state $W$, mentioned in Explanation \ref{entropie} in Section 2.2, can be written as $-\ln\langle w,w\rangle$, see also the final Remark in Section 2.3. Moreover, the \emph{marginal} densities of any {\scshape Wigner} density $w$ for momentum and position are the correct quantum probability densities, used
in the entropic \ini\ of Section 2.3:
\[\int_\R \frac{\dd q}{2\pi\hbar}\, w(p,q)= \langle p|W|p\rangle=\rho_P(p)\quad\text{and}\quad \int_\R \frac{\dd p}{2\pi\hbar}\, w(p,q)= \langle q|W|q\rangle=\rho_Q(q)\,. 
\]

\item Back to $w$ itself, {\scshape Hudson} has shown \cite{Hud1974} that a pure state $W=|\psi\rangle\langle\psi|$  has a positive {\scshape Wigner} density if, and only if, $\psi\in\LL^{2}(\R)$ is the exponential of a quadratic polynomial. For an example see Explanation \ref{uu} on the variance \ini\ in Section 2.2\,. {\scshape Cartwright} has shown \cite{Car1976} that any {\scshape Wigner} density becomes positive everywhere upon convolution with a {\scshape Gauss}ian probability density of product form on the phase space, confer the special $\widetilde{w}$ in Section 2.3, provided its momentum variance times its position variance is not smaller than $\hbar^2/4$. Her original proof is for pure states, but it easily extends to mixed states $W$ by their spectral resolutions. Moreover, if the variance product is equal to $\hbar^2/4$, then its factors can be chosen in such a way that the ``smoothed'' {\scshape Wigner} density becomes the {\scshape Husimi} density of $W$. Its values may be viewed as the expectation of $W$ with respect to (pure) canonical coherent states \cite{Bal1998,BZ2006,CFZ2019}. 
 \item {\scshape Moyal} bracket
  \begin{itemize}
   \item is a {\scshape Lie}--Product (without {\scshape Leibniz} rule).
   \item satisfies for ``small'' $\hbar\,$ informally: $\;\,\displaystyle \{a,b\} = [a,b] - \frac{\hbar^2}{24} a\, \mathcal{D}^3 \,b +{\cal O}(\hbar^4)$ and hence $\displaystyle \{a,b\} = [a,b],$ if $a$ or $b$ is at most quadratic in $p$ and $q$\,.
  \end{itemize}

 \item Instantaneous time-evolution (in the {\scshape Schrödinger} picture)
  \[   \frac{\dd}{\dd t} W_t = \left[W_t, H \right] \; \longleftrightarrow\; \frac{\dd}{\dd t} w_t = \left\{ w_t, h \right\}  \]

 \item Classical or quasi-classical limit (informally)
  \[   A \mapsto a \overset{\hbar \downarrow 0}{\longrightarrow} a_{\text{cl.}},\quad \{~,~\} \overset{\hbar \downarrow 0}{\longrightarrow} [~,~]  \]

 \item Since about 1960, {\scshape Weyl}--{\scshape Wigner} symbols and some of their ``relatives'' have found their way into the theory of \emph{pseudo-differential operators}, 
like $\exp \left(- \frac{\ii \hbar}{2} \mathcal{D} \right)$, and thereby gained great importance for the rigoros \emph{asymptotics of quasi-classical approximations} by {\scshape L. Nirenberg}, {\scshape J.J. Kohn}, {\scshape L. Hörmander}, {\scshape Yu.V. Egorov}, {\scshape M.A. Shubin}, {\scshape H. Widom}, {\scshape A.V. Sobolev}, and others. For the ``relatives'' see, for example, the concise overview \cite{FiLe1993} in terms of bi-orthogonal systems in the quantum {\scshape Liouville} space  and references therein.
 
 \item In Section 2.3 we have seen that the entropic indeterminacy inequality for momentum and position and, hence, also its weaker variance version, can be derived by {\scshape Fourier} analysis \cite{FoSi1997} without referring to {\scshape Hilbert}-space operators. Since this analysis also plays a major role in classical \emph{optics} and \emph{communications engineering}, there are analogs of the quantum indeterminacy inequalities in these fields for optical or other signals as functions of frequency or time (with frequency $\widehat{=}p$, time $\widehat{=}q$, and $2\pi\hbar\equiv1$). Similarly, in these fields one also makes use of analogs (and generalizations) of {\scshape Wigner} densities. 
 
\end{enumerate}

\section{Composite Systems, Reduced States, and Entanglement in Quantum Mechanics}
\setlength{\parindent}{1.8em}
Up to now, I have considered for simplicity only the system of a single particle moving along the {\scshape Euclid}ean line $\R$. The generalization to the $d$-dimensional {\scshape Euclid}ean space $\R^d$ with $d\in\{2,3,\dots\}$ is more or less straightforward and mainly a matter of notation. The position and the momentum both have $d$ components. The corresponding single-particle system has therefore the set-theoretic product $\Gamma_1\ceq\R^d\times\R^d\cong\R^{2d} $ as its phase space for the classical description and  $\H_1\ceq\LL^2 (\R^d)$ as its {\scshape Hilbert} space for the quantum description (in the {\scshape Schrödinger} realization). We may also consider a second particle in $\R^d$ with the same mass or not and being dynamically coupled to the first one or not. In any case, $\Gamma\ceq\R^{2d}\times\R^{2d}=\Gamma_1\times\Gamma_2$ is the total (or common) phase space and $\H\ceq\LL^2(\R^d\times\R^d)\cong \LL^2 (\R^d)\otimes \LL^2 (\R^d) = \H_1\otimes \H_2$ is the total {\scshape Hilbert} space of the (total) system composed of the two particles, which is a special case of a \emph{bipartite} system. We see that the role of the set-theoretic multiplication in the classical case, is taken over by the tensorial multiplication in the quantum case.

The extension from two to more particles should be obvious. Also the structural elements and notions compiled in the table of Section 2.1 have natural extensions for multi-particle systems in multi-dimensional space. However, two remarks are appropriate. First, the canonical relations hold as they stand (in the table), for each given particle with the \emph{same} component of its momentum and position. All other {\scshape Lie} products of two such components vanish. Second, if the particles are \emph{indistinguishable}, one should in the classical case divide all phase-space integrals by the factorial of the total number of particles. In the quantum case one should restrict the multi-particle {\scshape Hilbert} space either to its totally symmetric or anti-symmetric subspace, according to the {\scshape Bose--Einstein} resp. {\scshape Fermi--Dirac} statistics. The latter is the algebraic formulation of the {\scshape Pauli} exclusion rule \cite{Pa1925}.

In what follows, I will only consider the quantum case of a general bipartite system with two distinguishable subsystems \one\ and \two, possibly \emph{dynamically coupled}. That's general enough to exhibit certain quantum particularities. A prominent example is the hydrogen atom composed of a proton and an electron. Another one is the electron alone, but viewed as composed of a particle in $\R^3$ and its \emph{spin}\footnote{For details of this spin see, for example, Sections 4 and 5.1\,.}, an internal rotational degree of freedom.
\setlength{\parindent}{0em}

\medskip
Of basic interest are:\parbox[t]{0.77\textwidth}{
\vspace{-1.5em}\begin{itemize}\renewcommand{\labelitemi}{$\bullet$}
	\item The arena, observables, and states of the composite or \emph{total system} `` \one + \two'' composed of the subsystems. 
\item The states of the subsystems, given the state of the total system.
\end{itemize}}

\medskip{}
{\small\renewcommand{\arraystretch}{2}
\begin{tabular}{b{0.33\textwidth}cp{0.42\textwidth}}\hline
Arena of the total system: & $\H = \H_1 \otimes \H_2$&  (tensor product of {\scshape Hilbert} spaces)\\\hline
General observable: & $A$ & (self-adjoint operator on $\H$)\\\hline
\one-observable as an observable of the total system: &  $A_1\otimes\uu_2$ & (tensor product of operators)\\\hline
\two-observable as an observable of the total system: & $\uu_1\otimes A_2$& (tensor product of operators)\\\hline
\end{tabular}}

\bigskip
General observables $A$ of the total system are real (possibly infinite) linear combinations of \emph{product observables} 
$\quad A_1\otimes A_2= (A_1\otimes\uu_2)(\uu_1\otimes A_2)= (\uu_1\otimes A_2)(A_1\otimes\uu_2)$.

\subsection{Partial traces of an operator acting on a product {\scshape Hilbert} space}

\paragraph{Useful fact:}

If $(\varphi_n)_n$ resp. $(\psi_m)_m$ is an \emph{orthonormal basis} (ONB) in $\H_1$ resp. $\H_2$, then $(\varphi_n \otimes \psi_m)_{n,m}$ is an ONB in the tensor product $\H = \H_1 \otimes \H_2$. It is known as a \emph{product}-ONB and to be convenient for defining and calculating.\\

In particular, it can be used to write the \emph{trace} of an operator $A$ on $\H$ in the form 
\[
 \tr A = \sum_{n,m} \big\langle\varphi_n \otimes \psi_m\big| A(\varphi_n \otimes \psi_m)\big\rangle\,.
\]
\emph{Partial traces} of $A$ with respect to $\H_2$ or $\H_1$
\begin{itemize}
\item are operators on $\H_1$ resp. $\H_2$
\item are denoted by $\tr_2 A$ resp. $\tr_1 A$
\item can be defined through their (possibly infinite) matrix representations with respect to $(\varphi_n)_n$ resp. to $(\psi_m)_m$ as follows
\end{itemize}
\begin{center}
\boxedtext{\parbox{0.55\textwidth}{\vspace{-1em}
\begin{align*}
\big\langle\varphi_n\big| \tr_2 A\, \varphi_{n'}\big\rangle &\ceq \sum_m \big\langle\varphi_n \otimes \psi_m\big|A ( \varphi_{n'} \otimes \psi_m )\big\rangle\\
\big\langle\psi_m\big| \tr_1 A\, \psi_{m'}\big\rangle &\ceq \sum_n \big\langle\varphi_n \otimes \psi_m\big|A ( \varphi_{n} \otimes \psi_{m'} )\big\rangle
\end{align*}\vspace{-1.2em}
}}
\end{center}

\paragraph{Facts:}
\begin{itemize}\setlength\itemsep{0.25em}
\item Partial ``tracing'' is the analog of the ``classical'' integration with respect to a part of the phase-space variables.
\item The definitions of $\tr_2 A$ and $\tr_1 A$ are independent of the employed bases.
\item The partial trace of $A$ is self-adjoint (and positive) if $A$ is self-adjoint (and positive).
\item $\tr_2\big[ A (B_1 \otimes \uu_2 ) \big] = (\tr_2 A ) B_1, \quad \tr_1\big[ A(\uu_1 \otimes B_2) \big] = (\tr_1 A ) B_2$.
\item $\tr\big(\tr_2 A \big) = \tr\big(\tr_1 A \big) = \tr A\quad$ (in analogy to the {\scshape Fubini}--{\scshape Tonelli}\, theorem).
\end{itemize}

\paragraph{Reduced states}(or marginal states)

\medskip
of a state $W$ of the (bi-partite) total system are the two states
\[\boxedeqn{
 W_1 \ceq \tr_2 W, \quad W_2 \ceq \tr_1 W}
\]
on $\H_1$ resp. $\H_2$  of subsystem\, \one\, resp. \two. 

\medskip
Although they are operators, they are still often called ``reduced density matrices''. They correspond to the marginal densities in the \CM\ and \CPT.

\paragraph{Non-entangled and entangled states} (according to {\scshape Werner} \cite{Wer1989}):

\begin{itemize}\setlength\itemsep{0.25em}
\item A state $W$ on $\H=\H_1\otimes\H_2$ is called \emph{non-entangled} (\emph{separable} or \emph{classically correlated}) if it is for some $N\in\{1,2,3,\dots\}$ a convex combination of \emph{product states}\footnote{If $\dim \H=\infty$, then $W$ must be at least the (trace-norm) limit of the right-hand side as $N\to\infty$.}: 
\[\boxedeqn{W=\sum\limits_{n=1}^N p_n \,W_1^{(n)}\otimes W_2^{(n)}\,.}\]
 Here $W_1^{(n)}$ and $W_2^{(n)}$ are states on $\H_1$ resp. $\H_2$ and the numbers $p_n\geq 0$ sum up to $1$.

\item  A state $W$ on $\H_1\otimes \H_2$ is called \emph{entangled} if it is not non-entangled.

\end{itemize}

\paragraph{Remarks:}

\begin{itemize}\setlength\itemsep{0.25em}
\item A non-entangled state with $N=1$ is the product of its two reduced states
\[W=W_1\otimes W_2\,.\] 
It corresponds to stochastic independence in the \CM\ and \CPT\ and implies that the factors of all product observables $A_1\otimes A_2$ are uncorrelated: 
\[\big\langle A_1\otimes A_2\big\rangle=\tr(W_1\otimes W_2)( A_1\otimes A_2)=          \tr(W_1A_1\otimes  W_2A_2)=\big\langle A_1\rangle \big\langle A_2\big\rangle\,.\] 

\item A pure state is non-entangled if, and only if, it is, as above, the product of its two reduced states (and hence determined by a product vector).

\end{itemize}

\subsection{Two particularities of the reduction of quantum states}

in comparison to the reduction of classical states:

\begin{enumerate}[label=\arabic*)]
\item The reduction of a pure \QM\ state is in general mixed, because many pure states (of physical interest) are not a product and thus \emph{entangled pure states}.
\smallskip
\item Each mixed \QM\ state may be viewed as the reduction of a pure state on a ``larger''  {\scshape Hilbert} space. This goes under the name \emph{state purification}.
\end{enumerate}

\begin{proof} \begin{enumerate}[label=\arabic*)]
\item It is enough to consider a simple \QM\ system with product {\scshape Hilbert} space $\H_1\otimes \H_2$ from two-dimensional factors $\H_1 = \H_2 \cong\C^2$. This product is, for example, the arena of a system composed of \emph{two} spins with the same (intrinsic) ``main quantum number'' $\nicefrac{1}{2}$, but without the ``translational (or bosonic) degree of freedom'' considered in Section 2.
For a given ONB $\{\varphi_1,\varphi_2\}$ in $\H_1$ and an ONB $\{\psi_1, \psi_2\}$ in $ \H_2$, the two normalized linear combinations (``superpositions'')
\[\Phi^\pm \ceq \frac{1}{\sqrt{2}} ( \varphi_1 \otimes \psi_1  \pm \varphi_2 \otimes \psi_2 )\qquad(\Phi^\pm\in \H_1 \otimes \H_2)\]
define on $\H_1\otimes \H_2$ the two different states (confer the singlet state in Section 6)
\[W^\pm \ceq |\Phi^\pm\rangle\langle\Phi^\pm|\qquad (W^+W^-=\mathbb{0}) \,.\]
Both of them are pure and have analogous \emph{completely} (or ``chaotically'') mixed reduced states in the sense that
\[W_1^\pm = \frac{1}{2} \uu_1\quad\text{and} \quad W_2^\pm  = \frac{1}{2} \uu_2\,.\]
Since $\frac{1}{2}\uu_1\otimes\frac{1}{2}\uu_2\neq W^\pm $,  $W^+$ as well as $W^-$
is entangled.
\smallskip
\item If $W = \sum_n \rho_n |\psi_n\rangle\langle\psi_n|$ is the spectral resolution of the given state on some {\scshape Hilbert} space $\H$, then $\Phi \ceq\sum_n \sqrt{\rho_n}\psi_n \otimes \psi_n$ is a unit vector in $\H \otimes \H$. The pure state $|\Phi\rangle\langle\Phi|$ on $\H \otimes \H$ has the partial trace(s)\, $\tr_\H|\Phi\rangle\langle\Phi| = W$. \qed
\end{enumerate}
\end{proof}

I conjecture that both particularities are closely related to the quantum \emph{contextuality} explained in Section 5\,.

\subsection{Entropies, expectation values, and time-evolutions of subsystems}

\begin{enumerate}[label=\arabic*)]
\item Entropies
\smallskip
\begin{enumerate}[label=\alph*)]
\item \emph{total} entropy and \emph{reduced} entropies in the \QM\ case,
\begin{align*}
s\ceq - \tr W \ln W\,,\quad s_1 \ceq - \tr W_1 \ln W_1\,,\quad s_2 &\ceq - \tr W_2 \ln W_2\,,
\end{align*}
satisfy a triangle inequality in the sense of {\scshape Euclid}ean geometry
\[\boxedeqn{
 \left| s_1 - s_2 \right| \leq s \leq s_1 + s_2\,.}
\]

\paragraph{Visualization:}

\psset{unit=1cm}
\hfill\parbox{6cm}{
\begin{pspicture}(-0.5,-2)(5,0.5)
\psline(0,0)(4,0)
\psline(0,0)(5,-1.5)
\psline(4,0)(5,-1.5)
\rput(2.2,0.2){$s$}
\rput(2.6,-1.1){$s_1$}
\rput(4.8,-0.6){$s_2$} 
\end{pspicture}}

\vspace{-2em}
\begin{itemize}
\item plane of the triangle = drawing level
\item $s$, $s_1$, and $s_2$ correspond to the three side\\ lengths of the (here: obtused-angled) triangle.
\end{itemize}

\paragraph{Remarks:}
\smallskip
\begin{enumerate}[label=\roman*)]
\item The upper bound (known as\,\emph{subadditivity} and following from the quantum {\scshape Gibbs} inequality with $\widetilde W=W_1\otimes W_2$\,, see footnote 14\,.)
\begin{itemize}
\item holds also in the \CM\ and \CPT,
\item is interpretated as:\\ ``the information in $W$ is not smaller than the sum of those in $W_1$~and $W_2$'',
\item turns into an equality if, and only if, $W = W_1 \otimes W_2\,.$
\end{itemize}
\smallskip
\item The lower bound ({\scshape Araki--Lieb} \cite{AL1970})
\begin{itemize}
\item has no classical analog,
\item implies $s_1 = s_2$  if  $W$  is pure (see the geometric visualization for $s=0$),
\item yields, when combined with subadditivity, the implication\\
``$W_1$ \text{or} $W_2$\, \text{pure} $\quad \Longrightarrow\quad W = W_1 \otimes W_2$''.
\end{itemize}
\end{enumerate}
\smallskip
\item Entropy of \emph{entanglement} (or: relative entropy, mutual information)
\[\boxedeqn{\delta s\ceq s_1+s_2-s}\]
is a simple quantification of the entanglement between the two subsystems encoded in $W$. For mixed states $W$ this definition is in general too simple, in particular, because it may be ``spoiled'' by classical correlations.
\paragraph{Facts:}
\begin{enumerate}[label=\roman*)]
\medskip
\item $\delta s\geq0 $
\item $\delta s=0\,\,\,\, \quad\Longleftrightarrow\,\quad W=W_1\otimes W_2$
\item $W^2=W \quad\Longrightarrow\quad \delta s=2s_1=2s_2$
\end{enumerate}
\medskip
\item The two pure states $W^\pm$ of the previous subsection provide extreme examples with regard to entropy due to the following four facts:\\
\begin{align*}
s&=-\tr W^\pm\ln W^\pm=-1\ln 1 +3  (-0 \ln 0)=0 
\intertext{(minimal, \emph{since} the information about the total system is maximum),}
s_1&=-\tr W_1^\pm\ln W_1^\pm=-\frac{1}{2}\ln \frac{1}{2} -\frac{1}{2}\ln\frac{1}{2}=\ln 2\\
s_2&=-\tr W_2^\pm\ln W_2^\pm=\ln 2
\intertext{(maximum, \emph{although} the information about the total system is maximum!),}
\delta s&=s_1+s_2-s=2s_1=2s_2=2\ln 2
\intertext{(maximum).}
\end{align*}
\end{enumerate}

\item Expectation values of subsystem observables

\begin{center}
\boxedtext{\begin{minipage}{5.5cm}\vspace{-1em}\begin{align*}%
   \langle A_1 \otimes \uu_2\rangle &= \tr W(A_1 \otimes \uu_2) = \tr (W_1 A_1)\\
   \langle\uu_1 \otimes A_2\rangle &= \tr W(\uu_1 \otimes A_2) = \tr (W_2 A_2)\\
  \end{align*}\vspace{-3em}
 \end{minipage}
}\end{center}

The information encoded in $W_j$ suffices to answer all questions regarding system ~\kreis{$j$}\\ for $j \in \{1,2\}$. [analogous in the \CM\ and \CPT]
\medskip
\item Time-evolutions of subsytems
\begin{itemize}
\item They are simple in the case of \emph{dynamically independent} subsystems, that is, if the 
{\scshape Hamilton}ian of the total system ``\,\one + \two'' has the form
\[
 H =  H_1 \otimes \uu_2 + \uu_1 \otimes H_2
\]
without a coupling term. Then the subsystems evolve autonomously. More precisely, the
time-evolutions of the two reduced states of an arbitrary initial state of the total system are independent of each other and unitarily generated by $H_1$ resp. $H_2$.  As a consequence, all three entropies $s$, $s_1$, and $s_2$ do not change with time [analogously in the \CM].
\item They are more interesting (and complicated) in the general case of dynamically dependent (or coupled) subsystems, provided they cannot be decoupled by a suitable (canonical resp. unitary) transformation. Then the evolutions of the subsystems, inherited from the unitary evolution of the total system, are no longer unitary, intertwined with each other, and allow for time-dependent $s_1$ and $s_2$. This is nicely illustrated by the explicit example in Ch.7.5 of \cite{BC1981} for a \QM\ system with two spins, as mentioned in Section 3.2 and considered in Sections~5.2 and 6. In this example $s_1$ oscillates between its extreme values $0$ and $\ln 2$\,, corresponding to a pure resp. completely mixed (reduced) state of subsystem~\one.
\end{itemize}
\end{enumerate}

\section{Quantum Events and the Theorem of {\scshape Gleason}}

\begin{itemize}
 \item The set $\mathcal{P}(\H)$ of \emph{all} projections $E,F,G,\dots$ on a separable {\scshape Hilbert} space $\H$ forms \cite{Ja1977,Hug1989,V2007}
an \emph{orthomodular} lattice or \emph{partial} {\scshape Boole}an algebra with respect to the three operations:
 
\medskip
\begin{tabular}{rllll}
i)	&$\displaystyle E^\bot \ceq \uu - E$ &projects on &$(E\H)^\bot$ &(orthog. complement)\\
ii)	&$\displaystyle E \wedge F \ceq \lim_{n \rightarrow \infty} E(FE)^n$ &projects on &$ E\H \cap F\H$ &(intersection)\footnotemark\\
iii)	&$\displaystyle E\vee F \ceq \left(E^\bot \wedge F^\bot \right)^\bot$ &projects on
 &$ \overline{E\H + F\H}$ &(closed linear sum).
\end{tabular}

\footnotetext{If $\dim\H=\infty$, the convergence is understood in the strong operator-topology, see p.71 in \cite{BEH2008}.}
\bigskip
If $\dim\H\geq 3$, then $\mathcal{P}(\H)$ is not a {\scshape Boole}an algebra.

\bigskip
Geometric visualization of ii):

\parbox{7cm}{
\psset{unit=1cm}
\begin{pspicture}*[algebraic](-1,-1)(5.7,4)
\psset{arrowscale=1.6}
\rput(5.3,0){$E\mathcal{H}$}
\rput(3.2,3.2){$F\mathcal{H}$}
\psline[linewidth=0.5pt](-0.75,0)(5,0)
\psline[linewidth=0.5pt](-0.5,-0.5)(3,3)

\psline[linewidth=0.5pt,linestyle=dashed,ArrowInside=->,ArrowInsideNo=1,arrowsize=2pt,linecolor=black!75](4,1)(4,0)
\psline[linewidth=0.5pt,linestyle=dashed,ArrowInside=->,ArrowInsideNo=1,arrowsize=2pt,linecolor=black!75](4,0)(2,2)
\psline[linewidth=0.5pt,linestyle=dashed,ArrowInside=->,ArrowInsideNo=1,arrowsize=2pt,linecolor=black!75](2,2)(2,0)
\psline[linewidth=0.5pt,linestyle=dashed,ArrowInside=->,ArrowInsideNo=1,arrowsize=2pt,linecolor=black!75](2,0)(1,1)
\psline[linewidth=0.5pt,linestyle=dashed,ArrowInside=->,ArrowInsideNo=1,arrowsize=2pt,linecolor=black!75](1,1)(1,0)
\psline[linewidth=0.5pt,linestyle=dashed,ArrowInside=->,ArrowInsideNo=1,arrowsize=2pt,linecolor=black!75](1,0)(0.5,0.5)
\psline[linewidth=0.5pt,linestyle=dashed,ArrowInside=->,ArrowInsideNo=1,arrowsize=2pt,linecolor=black!75](0.5,0.5)(0.5,0)
\psline[linewidth=0.7pt,linecolor=blue]{->}(0,0)(4,1)
\psline[linewidth=0.7pt,linecolor=blue]{->}(0,0)(4,0)
\psline[linewidth=0.7pt,linecolor=blue]{->}(0,0)(2,0)
\psline[linewidth=0.7pt,linecolor=blue]{->}(0,0)(1,0)
\psline[linewidth=0.7pt,linecolor=blue]{->}(0,0)(0.5,0)
\psline[linewidth=0.7pt,linecolor=blue]{->}(0,0)(2,2)
\psline[linewidth=0.7pt,linecolor=blue]{->}(0,0)(1,1)
\psline[linewidth=0.7pt,linecolor=blue]{->}(0,0)(0.5,0.5)
\rput(4.2,1.1){\textcolor{blue}{$\psi$}}
\rput(4,-0.3){$E\textcolor{blue}{\psi}$}
\rput(1.65,2.2){$FE\textcolor{blue}{\psi}$}
\rput(2,-0.3){$EFE\textcolor{blue}{\psi}$}
\rput(0.4,1.2){$(FE)^2\textcolor{blue}{\psi}$}
\psdot(0,0)
\rput(0.15,-0.3){$\{{\textcolor{blue}0}\}$}
\end{pspicture}
}
\parbox{4.5cm}{\small Alternating action of two projections\\ $E\neq F$ (on $\H=\R^2=$ drawing level) with $EF\neq FE$ and $E\wedge F=\mathbb{0}$.\newline 
[For $\dim\H=2$ one has:\\ $``EF=FE\Longleftrightarrow EF=\mathbb{0}\text{''} \,.$]
}

\medskip
\item Simplifications for \emph{compatible} events, that is, for pairwise \emph{commuting} projections ($EF=FE$ etc.):
\[ E \wedge F = EF, \quad E\vee F = E+F - EF\,.\]
These imply the {\scshape Boole}an distributive laws:
  \begin{align*}
   E \wedge (F \vee G) &= (E \wedge F) \vee (E \wedge G)\\
   E \vee (F \wedge G) &= (E \vee F) \wedge (E \vee G)\,.
  \end{align*}
 
\item Each subset of pairwise commuting projections in $\mathcal{P}(\H)$ with $\uu$ and $\mathbb{0}$ is isomorphic to the {\scshape Boole}an algebra of subsets of some basic set with the set-theoretic operations complement, intersection, and union.
\end{itemize}

\begin{center}
 \boxedtext{
 \begin{minipage}{\textwidth-11pt}
\begin{theorem*} 

\medskip
If $W$ is a state on $\H$, then the mapping $\mu : {\mathcal P}(\H)\to\R$, defined by
$E \mapsto \mu(E) \ceq \tr(WE)$, is a \emph{probability measure} on $\mathcal{P}(\H)$ in 
the sense that it has the following three properties:
\begin{enumerate}[label=\alph*)]
 \item\label{alpha} $\displaystyle 0 \leq \mu(E) \leq 1$
 \smallskip
 \item\label{beta} $\displaystyle \mu(\mathbb{0})=0,\; \mu(\uu)=1$
 \smallskip
 \item\label{gamma} $\displaystyle \mu\Big(\bigvee_n E_n \Big) = \sum_n\mu(E_n)$\\ for any sequence $(E_n)$ of pairwise \emph{disjoint} and pairwise \emph{commuting} projections.
\end{enumerate} 
\end{theorem*}
 \end{minipage}}
\end{center}

\paragraph{Explanation:}

\medskip
 If $EF=FE$, then the following equivalences hold (see above):
 \[
   \text{``}\,E\,\, \text{and}\,\,  F\,\, \text{are \emph{disjoint}} : \Longleftrightarrow\ E\wedge F = \mathbb{0} \, \Longleftrightarrow\, EF=\mathbb{0}  \Longleftrightarrow\, E \vee F = E+F\,
 \,\text{''}. \]
Property \ref{gamma} in the theorem can therefore be recognized as the  \emph{countable additivity} of $\mu$\,:\\ $\mu(\sum_n E_n) = \sum_n \mu(E_n)$ for any sequence $(E_n)$ of pairwise \emph{orthogonal} projections, that is, $E_n E_m= \delta_{nm} E_n\,.$
The proof of the theorem is hence obvious if $\dim\H<\infty$. Even the proof of countable additivity in the case $\dim\H=\infty\,$ is not hard, see Theorem A11 in \cite{BC1981}. In contrast, the following \emph{reverse} statement of the theorem is profound and its proof therefore arduous \cite{Gl1957,V2007}. A simplified proof is given in \cite{CKM1985}, see also \cite{Hug1989}.

\begin{center}
 \boxedtext{
 \begin{minipage}{\textwidth-11pt}
\begin{theorem*}({\scshape Gleason} 1957)
 
\medskip
If $\dim \H \geq 3$, then there exists for each probability measure $\mu$ on $\mathcal{P}(\H)$ a unique state $W$ on $\H$ such that $\mu(E) = \tr(WE)$ for all $E \in \mathcal{P}(\H)$.{}
\end{theorem*}
 \end{minipage}}
\end{center}

\paragraph{Remarks:}

 \begin{itemize}
 \item Each ONB in $\H$ provides an example of a sequence of pairwise orthogonal (one-dimensional) projections. The depth (and surprise?) of the result is related to the fact that  additivity is only required for commuting projections. 
 \item Given the notion of a \emph{quantum} event, the result is fundamental because it justifies the notion of a \emph{quantum} state\footnote{for systems with \emph{finitely} many degrees of freedom} and also, to some extent, the {\scshape Born--von Neumann} probability postulate of Explanation \ref{wsk} in Section 2.2\,.
\item The \emph{quantum} (or \emph {non-commutative}) probability theory (QPT) with the lattice/algebra of projections is more general than the classical one (\CPT), but reduces to \CPT\ for any ({\scshape Boole}an) \emph{sublattice} of $\mathcal{P}(\H)$ containing only pairwise \emph{commuting} projections including $\mathbb{0}$ and $\uu$, see {\scshape Faris} in \cite{W1995} and \cite{Ja1977,Heg1985,Stre2000,Ac2010}.
 
\paragraph{Analogy:}
QPT extends \CPT\ in a similar way to how non-{\scshape Euclid}ean geometry extends the {\scshape Euclid}ean one \cite{Ja1977,Stre2007,Ac2010}. The outcomes of experiments decide which theory is better suited to model aspects of Nature.
\end{itemize}

For $\dim \H = 2$ there exist probability measures on $\mathcal{P}(\H)$ that differ from those offered by the {\scshape Gleason} theorem for $\dim\H\geq 3$. This fact can be seen as follows.

\medskip
\begin{enumerate}[label=\alph*)]

\item Explicit description of all elements of $\mathcal{P}(\C^2)$:

\medskip
A \QM\ system with a two-dimensional {\scshape Hilbert} space $\H\cong\C^2$ is the simplest non-trivial one. It is the arena for a single spin with (intrinsic) ``main quantum number'' $\nicefrac{1}{2}$. Up to a factor $\hbar/2$, the corresponding vector operator is $\vv{S} \ceq (S^{x}, S^{y}, S^{z})$ and represented in the eigenbase of $S^{z}$ by the triple of $2\times2$ ({\scshape Dirac}--){\scshape Pauli} matrices according to:
 \[S^{x}\,\widehat{=}\begin{pmatrix} 0 &1\\ 1 &0  \end{pmatrix},\,\; S^{y}\,\widehat{=}\begin{pmatrix} 0 &-\ii\\ \ii &0 \end{pmatrix},\,\; S^{z}\,\widehat{=} \begin{pmatrix} 1 &0\\ 0 &-1 \end{pmatrix}
 \,.\]

They are self-adjoint, their squares are equal to the ${2\times 2}$ unit matrix representing the unit operator $\uu$, and they imply the identities
 $S^{x}S^{y}=\ii S^{z}$, $S^{y}S^{z}=\ii S^{x}$, and $S^{z}S^{x}=\ii S^{y}$\,.
 \newpage

The lattice/algebra $\mathcal{P}(\C^2)$ consists of the following projections:\\

\arrayrulecolor{red}
\begin{tabular}{p{0.28\textwidth}|p{0.36\textwidth}|}\cline{2-2}&\\
one 0-dimensional & $\mathbb{0}$ \\&\\
one 2-dimensional & $\uu$ \\&\\
a two-parameter \newline family of 1-dimensional &  $\displaystyle E\ceq \frac{1}{2} (\uu + \vv{e} \boldsymbol{\cdot} \vv{S})$,\newline $\vv{e} \in\R^3$ with $|\vv{e}| = 1$ \medskip\\\cline{2-2}
\end{tabular}
\arrayrulecolor{black}

\begin{tabular}{lp{0.68\textwidth}}
Implication: &$(\vv{e}\boldsymbol{\cdot} \vv{S})E=E$\medskip\\
Interpretation:& $E$ is a pure eigenstate of the spin component\newline
$\vv{e}\boldsymbol{\cdot}\vv{S}\ceq e^{x}S^{x} + e^{y}S^{y} + e^{z}S^{z}$ along the $\vv{e}$ direction\newline with eigenvalue $1$.
\end{tabular}

\bigskip
\item Explicit description of all probability measures on $\mathcal{P}(\C^2)$:
\medskip
\[\boxedeqn{\mu(0) \ceq 0,\;\, \mu(\uu) \ceq 1,\;\, \mu(E) \ceq \frac{1}{2} \left(1 + m(\vv{e})\right)}\]
Here $m$ is an arbitrary mapping of the surface $\mathbb S^2$ of the closed unit ball in $ \R^3 $ to the interval ${[-1,1]}$  with\, $m(-\vv{e}) = - m(\vv{e})$\, (to ensure that $\mu(E) + \mu(E^\bot) = 1$).
\newline Simplest example:\; $m(\vv{e})=0$\, for all unit vectors $\vv{e}$.\\
\end{enumerate}
A comparison with the three formulas $\,\tr(W\mathbb{0})=0$, $\,\tr(W\uu)=1\,$, and\\ $\,\tr(WE)=\frac{1}{2}\big(1+\vv{e}\boldsymbol{\cdot} \tr(W\vv{S})\big)\,$ now yields the\\ \\
\smallskip
\textbf{Equivalence}:
\newline
\boxedtext{
\begin{tabular}{p{0.27\textwidth}p{0.05\textwidth}p{0.46\textwidth}}
$\mu$ is induced by a state $W$ on $\C^2$ &$\Longleftrightarrow$ &There is $\vv{p} \in \R^3$ with $|\vv{p}~| \leq 1$ such that \newline $m (\vv{e}) =\vv{p}\boldsymbol{\cdot} \vv{e}$ for all $\vv{e}\in\R^3$ with $|\vv{e}| = 1$
\end{tabular}}
\smallskip
\begin{proof} Choose $\vv{p} = \tr(W \vv{S})$ and conversely $W = \frac{1}{2}(\uu   + \vv{p} \boldsymbol{\cdot} \vv{S})$\,.\qed\end{proof}
The above claimed existence of probability measures \emph{not} arising from a state $W$ on $\C^2$, is now easy to see.
Here are two examples (with some $\vv{\lambda}\in\R^3$): 
\begin{enumerate}[label=\roman*)]
 \item $\displaystyle m(\vv{e}) = \sin(\vv{\lambda}\boldsymbol{\cdot} \vv{e}),\qquad \sin(\xi)\ceq\frac{1}{2\ii}\big(\textrm{e}^{\ii\xi}-\textrm{e}^{-\ii\xi}\big)\quad$ for $\quad\xi\in\R$
\bigskip
 \item $\displaystyle m(\vv{e}) = \sgn(\vv{\lambda}\boldsymbol{\cdot} \vv{e}),\;\;\quad \sgn(\xi)\ceq -1,\;  0,\;\text{or}\; 1 \quad$ if $\quad \xi<0,\;  =0, \;\text{or} \; >0\,.$ 
\end{enumerate}
The resulting measures $\mu$ do not seem to have any physical relevance, but some of their ``relatives'' play a role in classical models with hidden variables for \QM\ systems with $\dim {\cal H}=2$, see Section 5.1\,.

\section{On the Realistic Interpretability of Quantum Mechanics}

Edward {\scshape Nelson} (1932--2014)\;\; (in \cite[p. 230]{Far2006}):\\

``\emph{The problem of finding a realistic interpretation of quantum mechanics is, in my view, as unresolved as it was in the 1920s}.'' 

\paragraph{Findings}\,

\medskip
The {\scshape Hilbert}-space formalism of quantum mechanics (QM) by {\scshape Dirac} and {\scshape von Neumann} together with its pragmatic and minimal(ist) \emph{statistical interpretation} -- as summarized above -- has developed over the decades into the most successful theory of physics.
 
\paragraph{Discomfort of classical realists}

(like {\scshape Planck}, {\scshape Einstein}, {\scshape deBroglie}, {\scshape Schrödinger}, D. {\scshape Bohm}, {\scshape Bell}, {\scshape 't Hooft}\footnote{see \cite{'t Hoo2024}},\dots) 

\medskip
Even in pure states, ``most'' observables do not have a specific value, but are \emph{objectively indeterminate}. Values only arise as factual outcomes of measurements with an \emph{intrinsic} probability $\langle W, E\rangle$, as briefly described in Explanation\,\ref{varianz} in Section\,2.2 about the ``state collapse''. This contrasts with a sentence by {\scshape Einstein} to {\scshape Born} in December 1926, often shortened but pithily quoted as ``Gott würfelt nicht.''\footnote{In English: ``God does not play dice.''} Even to this day, no proposal at understanding the quantum indeterminacy or accepting it as a feature of (a veiled?) reality \cite{d'Esp2003} has met with general approval. This is, of course, related to the infamous \emph{measurement problem}. For some discussions of the latter see \cite{Zeh1970,Mit1998,Ho2001,Zeh2007,BLPY2016,Hay2017}.
\medskip

\paragraph{Hope for a way out by cryptodeterminism} (Models with hidden variables (\HV))

\medskip
The indeterminacy, reflected by a strictly positive variance, $\sigma_A^2 > 0$\,, in non-eigenstates of an observable $A$ can be explained by ``embedding'' the \QM\ into a more fundamental probabilistic theory, which is \emph{realistically} (or ``classically'') interpretable in the following sense:
\begin{enumerate}[label = \roman*)]

\item Each observable $A$ of an \emph{individual} \QM\ system has in each (pure) state $W$ an \emph{objectively definite} (eigen)value, pre-existing before any measurement.
\smallskip
 \item The \QM\ probabilities only express that the distribution of these values is in general not concentrated at a single point, unknown, and empirically in-accessable, that is,  \emph{hidden}.  

\end{enumerate}

\medskip
In short: \hfill\boxedtext{\emph{Subjective ignorance} instead of \emph{objective indeterminacy}}\hfill\,

\bigskip
[somewhat similar to \emph{statistical} \CM].

\subsection{Basic idea behind most models with hidden variables}

\begin{itemize}
 \item Starting point:
  \QM\ system with {\scshape Hilbert} space $\H$, observables $A: \H \rightarrow \H$, and a state $W: \H \rightarrow \H$ mostly assumed to be pure.

  \item Classical \emph{probability space} \cite{Bau1996}:
 $\big(\Omega, \mathcal{A}(\Omega),\P\big)$, that is, a triple consisting of the set $\Omega$ of the \emph{hidden variables} $\omega$, a sigma-algebra $\mathcal{A}(\Omega)$ of subsets of $\Omega$, and a probability measure $\P$ on $\Omega$, more precisely, on $\mathcal{A}(\Omega)$.
\item Desired identity:
 \[\boxedeqn{\int_\Omega \P(\dd\omega)\, a_W(\omega) = \tr(WA)}\]
 for a suitable  \emph{value-function}  $a_W :\Omega\rightarrow \R$ assigned to the pair $(A,W)$. In \CPT\ such a function is usually called a random variable.
\end{itemize}

\paragraph{Remarks:}

 \begin{itemize}
 \item The \emph {hidden variables} $\omega\in\Omega$ discriminate between the individual systems (realizations).
\item The probability measure $\P$ also may depend on $W$, or $W$ may only be ``transferred‘‘ from $a_W$ to $\P$. The {\scshape Weyl}--{\scshape Wigner} mapping of Section 2.5 provides something similar to the latter case with the phase space $\Gamma$ playing the role of $\Omega$, the {\scshape Wigner} density $w$ that of $\P_W$, and the symbol $a$ of $A$ that of $a_W$. However, one serious problem with this ``could-be'' \HV-model for the canonical \QM\ is that $w$ is in general not positive everywhere. Another one is that the symbol of $A^2$ is in general not $a^2$ etc.

\end{itemize}  

\emph{Interpretation:} 

\medskip
While the description by a (pure) \QM\ state $W(=W^2)$ may not provide a ``complete picture'' of the physical reality on its own, the pair $(W, \omega)$, the so-called \emph{microstate}, does so in the sense of the \emph{assignment}:
\[ \big(A, (W, \omega) \big) \mapsto a_W(\omega) = \text{some (eigen)value of the observable $A$ in the microstate $(W,\omega)$}\,.\]

\paragraph{Model with hidden variables in the case of a two-dimensional {\scshape Hilbert} space:}

\begin{itemize}
 \item {\scshape Hilbert} space $\H = \C^2$ associated with a single spin $\nicefrac{1}{2}$ as in Section\,4.{}
 \smallskip
 \item Useful relation for the product of two general spin-operator components:
  \[\boxedeqn{
 \big(\vv{a}\cdot\vv{S}\big)\big(\vv{b}\cdot\vv{S}\big)=\vv{a}\cdot\vv{b}\uu+\ii\big(\vv{a}\times\vv{b}\big)\cdot\vv{S}
 }\]
 for \emph{any} vectors $\vv{a},\vv{b}\in\R^{3}$ in terms of their dot product and cross product.
\smallskip  
  \item Each \QM\ observable $A$ on $\C^2$ is a linear combination
   \[\boxedeqn{
    A = \frac{1}{2}\big(a_0 \uu + \vv{a}\boldsymbol{\cdot} \vv{S}\big)}
    \] 
     with $a_0 \ceq \tr A \in\R$ and $\vv{a} \ceq \tr(A \vv{S})\in\R^3$,
    as a generalization of $E=\frac{1}{2}(\uu + \vv{e}\boldsymbol{\cdot} \vv{S})$ in
    the set $\mathcal{P}(\C^2)$ explicitly described in Section 4.
\smallskip
\item Equivalence (if $\vv{a}\neq\vv{0}\neq\vv{b}$):
\[\boxedeqn{
 AB = BA \; \Leftrightarrow \; \vv{a}\text{ and }\vv{b}\,\text{are parallel or anti-parallel (that is, \emph{collinear})}\,,} \]
by the commutation relation $AB-BA=2 \ii\big(\vv{a}\times\vv{b}\big)\cdot\vv{S}$ from the above ``useful relation''.
\smallskip
\item Each \QM\ state $W$ on $\C^2$ can be written similarly to observables (recall also Section~4):
  \[\boxedeqn{W = \frac{1}{2} \big( \uu + \vv{p}\boldsymbol{\cdot} \vv{S}\big)\,.}  \]
The vector $\vv{p} \ceq \tr(W\vv{S})\in\R^3$ is known as the \emph{polarization} of $W$ and obeys $|\vv{p}\,|\leq 1$. The pure states on $\C^2$ (or \emph{quantum bits} of quantum information theory) are exactly the \emph{totally} polarized ones: 
  \[\boxedeqn{ W = W^2  \; \Leftrightarrow \; |\vv{p}\,| = 1\,. }\]
 
Conclusion:

\smallskip
\boxedtext{\parbox{0.93\textwidth}{Each state $W$ corresponds exactly to one point of the closed unit ball in $\R^3$ (called {\scshape Bloch} ball in this setting). The pure states correspond to the points on its surface ${\mathbb S}^2$.}}
\medskip
\item Expectation values now take the form: 
  \[\boxedeqn{
   \tr(WA) = \frac{1}{2}\big(a_0 + \vv{p}\boldsymbol{\cdot}\vv{a}\big)\,.}
  \]
\end{itemize}
With the choices 
\begin{center}\renewcommand{\arraystretch}{1.5}
 \boxedtext{\begin{tabular}{l}
$\Omega={\mathbb S}^2 \ceq$ unit sphere in $ \R^3 $\\
${\cal A}(\Omega)={\cal B}(\mathbb{S}^2)\ceq$ sigma-algebra of all {\scshape Borel} subsets of ${\mathbb S}^2$\\
 $\P =$ uniform distribution on $\Omega$\\
  $\displaystyle a_W (\vv{\omega}) \ceq\frac{1}{2}\big( a_0 + |\vv{a}|
  \sgn\left((\vv{p} + \vv{\omega})\boldsymbol{\cdot} \vv{a} \right)\big),\qquad \vv{\omega}\in\Omega$
 \end{tabular}}
\end{center}

one now indeed gets, by explicit integration in spherical coordinates, the desired identity
\[
	\int_\Omega \P(\dd \vv{\omega})\, a_W(\vv{\omega})
	= \frac{1}{4\pi}\int_{\R^3}\!\dd^3\!\omega \,\,\delta\big(|\vv{\omega}| - 1\big)\,a_W(\vv{\omega}) = \frac{1}{2}\big(a_0+\vv{p}\boldsymbol{\cdot}\vv{a}\big)=\tr(WA)\,.{}
\]
\textbf{Facts}\\ on the above non-linear assignment\, $A\mapsto a_W$\; (for fixed $W$, equivalently, fixed $\vv{p}$)\,:\\
\vspace{-0.3em}
\begin{itemize}
\item the values $a_W(\vv{\omega})$ of $a_W$ coincide for $\P$-almost
all $\vv{\omega}$ with one of the two eigenvalues\\ $(a_0 \pm|\vv{a}|)/2$ of the observable $A\;\leftrightarrow\; (a_0, \vv{a})$. This is known as \emph{spectral consistency}.\\ In particular, 
\[e_W(\vv{\omega})=\frac{1}{2}\Big(1+\sgn\big((\vv{p}+\vv{\omega})\boldsymbol\cdot\vv{e}\big)\Big)\] 
is the probability $\mu(E)$ of the 1-dimensional projection $E\;\leftrightarrow\; (1,\vv{e})$, resulting from the example~ii) at the end of Section~4 by choosing $\vv\lambda=\vv{p}+\vv{\omega}$.
 \item the (eigen)value $c_W(\vv{\omega})$ of the sum $C\ceq A + B$ is \emph{additively consistent} in the sense that $c_W(\vv{\omega}) = a_W(\vv{\omega}) + b_W(\vv{\omega})$ if \boxedtext{$AB=BA$\,.} (Recall that $\vv{a}$ and $\vv{b}$ are then collinear.)
 
\item the (eigen)value $d_W(\vv{\omega})$ of the product $D\ceq AB$ is \emph{multiplicatively consistent} in the sense that $d_W(\vv{\omega}) = a_W(\vv{\omega}) b_W(\vv{\omega})$ if  \boxedtext{$AB=BA$\,.} In particular, the microstate $(W,\vv{\omega})$ is universally fluctuation-free because $d_W(\vv{\omega}) = \big(a_W(\vv{\omega})\big)^2$ for $A=B$.

\item the assignment is \emph{injective} (or one-to-one) in the sense that\;``$a_W(\vv{\omega})= b_W(\vv{\omega})$ for\\ $\P$-almost all $ \vv{\omega}$\,'' implies $A=B$\;.
\end{itemize}\medskip

Remarks:
\begin{itemize}
\item Such \HV-models for $\dim \H = 2$ go back to a publication of {\scshape J.S. Bell}  in 1966 (submitted in 1964), see the first paper re-printed in \cite{Bel2004}.
\item For $\dim \H \geq 3$ such \HV-models do not exist (for \emph{all} observables), because then there may emerge a conflict between the above consistencies and the injectivity. This will be explained in the next subsection.
\end{itemize}

\subsection{Impossibility theorem of {\scshape Bell} and {\scshape Kochen}--{\scshape Specker}}

The following theorem refers to a \emph{single} individual \QM\ system and does not provide a statement on the statistics of quantum measurement outcomes. Accordingly, it does not explicitly refer to a state (before measurement). Instead it analyzes assignments $A\mapsto a$ to (pre-existing) real (eigen)values for a putative \emph{realistic interpretation} from a more fundamental point of view, regardless whether they depend on a fixed state $W$ or not. No averaging takes place. For the original works see the first paper re-printed in \cite{Bel2004} and \cite{KS1967}. 

\begin{center}
\smallskip
 \boxedtext{
 \begin{minipage}{\textwidth-11pt}
\begin{theorem*} ({\scshape Bell} 1966, {\scshape Kochen}--{\scshape Specker} 1967)

\medskip
Let $\dim \H \geq 3$ and $\mathcal{O(\H)}$ be the set of self-adjoint (and bounded) \emph{operators} $A:\H\rightarrow\H$\,. Moreover, let $\big(\Omega, \A(\Omega)\big)$ be a measurable space [$=$ \quotes{phase space} $=$ set of hidden variables] and $\mathcal{F}(\Omega)$\, be the set of $\big(\A(\Omega), \B(\R)\big)$-measurable \emph{functions} $a:\Omega\rightarrow \R,\; \omega \mapsto a(\omega) $ [$=$ value-functions and possible random variables].\footnote{The notion ``measurable'' is meant here in the sense of mathematical measure- and integration theory. The function $a:\Omega\rightarrow \R$ is defined to be $\big(\A(\Omega), \B(\R)\big)$-\emph{measurable} if the pre-image $a^{-1}(I)\ceq\{\omega\in\Omega: a(\omega)\in I\}$ of any {\scshape Borel} set $I\subseteq\R$ belongs to the sigma-algebra $\A(\Omega)$ of suitable subsets of $\Omega$.}\\
\newline
Then there exists \emph{no} mapping [$=$ ``classical representation'']
\[
 \Theta:\mathcal{O(\H)} \rightarrow \mathcal{F}(\Omega),\qquad A\mapsto a\ceq \Theta (A)
\]
with the following two properties:
\begin{enumerate}[label=\arabic*)]
 \item $\Theta$ is \emph{injective} (or one-to-one) in the sense of the usual implication\\ 
  \[\text{``}\; a(\omega)=b(\omega)\quad\text{for all}\;\; \omega\in\Omega \; \Longrightarrow \; A=B\; \text{''}   \,, \]
\item $\Theta$ is \emph{functionally consistent} in the sense of the equality\\
  \[\Theta\big(f(A)\big)(\omega) = f\big(\Theta(A)\big)(\omega)\quad\text{for all}\;\; \omega\in\Omega \]
  and any {\scshape Borel} function $f:\; \R \rightarrow \R\,.$
\end{enumerate}
\end{theorem*} \end{minipage}}
\end{center}

\paragraph{Comments}
\begin{itemize}
\item
The image $f(A)$ of the operator $A$ under a (non-polynomial) function $f$ is defined, as usual, by the spectral resolution of $A$ according to the \emph{functional calculus}, see \cite{BEH2008}. In fact, I have tacitly used this already above. Examples:\, $\chi_I(A),\,
 \textrm{e}^{-\beta H},\, W\ln W,\, W^{1/2}$\,.
\item
\emph{Functional consistency} is the physical one of the two properties required. It simply ensures that the value $f(a(\omega))$ of the composed function $f(a)\ceq f\circ a$ at $\omega\in\Omega$ agrees with $f(\alpha)$ for some eigenvalue $\alpha$ of $A$. Example: If a measurement reveals an eigenvalue of $A^2$, then it should be the square of some eigenvalue of $A$.
\newpage
Functional consistency and injectivity are illustrated by the following diagram:

\begin{center}
\psset{unit=1cm}
\begin{pspicture}*[algebraic](-0.6,-0.6)(2.6,2.5)
\psset{arrowscale=1.6}
\rput(2,2){$f(A)$}
\rput(0,2){$A$}
\rput(0,0){$a$}
\rput(2,0){$f(a)$}
\rput(1,-0.3){$f$}
\rput(1,2.3){$f$}
\rput(-0.3,1){$\Theta$}
\rput(2.3,1){$\Theta$}
\psline[linewidth=0.5pt]{|->}(0.3,2)(1.6,2)
\psline[linewidth=0.5pt]{|->}(0.3,0)(1.6,0)
\psline[linewidth=0.5pt]{<->}(2,0.3)(2,1.6)
\psline[linewidth=0.5pt]{<->}(0,0.3)(0,1.6)
\end{pspicture}
\end{center}
 
\item
The formulation of the theorem, given above, may appear slightly complicated and tend to veil its central message. If so, in a first reading one may ignore the ``sigma-algebras'' $\A(\Omega)$ and $\B(\R)$ and understand ``measurable function'' and ``{\scshape Borel} function'' simply as ``function''. The remarkable result is that any possible mapping $\Theta$ assigning to each self-adjoint operator $A$ a ``corresponding'' real-valued function $a$ must violate either the injectivity or the functional consistency. Most physicists in search of a realistic interpretation will not renounce the latter for good physical reasons, see above. As it turns out, see below, they therefore have to accept that one and the same quantum observable may be assigned to different classical representations depending on the specific physical \emph{context} in which other commuting (!) observables are considered for (simultaneous) measurements. This is, for example, the case in the well-confirmed {\scshape Bell}-test experiments, see Section 6. Injective representations, that is,``universal'' or \emph{non-contextual} ones are mostly rejected because they lack full functional consistency and therefore cannot fully describe the statistics of quantum measurement outcomes. This is often summarized as: The B--KS theorem rules out non-contextual \HV-models. For further informations and discussions see \cite{Hug1989,Mer1993,I1995,Per2002,Stre2007,Cab2021,BCGKL2022}.
\end{itemize}

\paragraph{Some consequences of injectivity and functional consistency}\,

\begin{enumerate}[label=\roman*)]
\item\label{folg_a} Finitely additive consistency: $\quad\displaystyle \Theta\Big({\sum_{j=1}^n A_j}\Big)=  \sum_{j=1}^n \Theta (A_j)$, \\
  for pairwise \emph {commuting} $A_1, \dots, A_n \in \mathcal{O(\H)}$
\item\label{folg_b} Finitely multiplicative consistency: $\quad\displaystyle \Theta\Big({\prod_{j=1}^n A_j}\Big)  = \prod_{j=1}^n \Theta({A_j})
 $,\\ for pairwise \emph{commuting} $A_1, \dots, A_n \in \mathcal{O(\H)}$
 \item\label{folg_c} Homogeneous consistency: $\quad\displaystyle \Theta(\lambda A) = \lambda  \Theta(A)$,\; $\lambda \in \R$
 \item\label{folg_d} Spectral consistency: $\quad\displaystyle \Theta(A)(\Omega)\equiv a(\Omega) \subseteq \spec A$ ($\ceq$ spectrum of $A$)
\end{enumerate}

All of them follow from the

\begin{lemma*}({\scshape von Neumann 1931})\\
For any collection of pairwise \emph{commuting} $A_1,\dots, A_n\in \mathcal{O(\H)}$\, there exist\,a bounded $X\in\mathcal{O(\H)}$\, 
and a collection of \, {\scshape Borel} functions $f_1, \dots, f_n :\; \R \rightarrow \R$\, such that \;$A_j = f_j(X)$\, for all  $j\in \{1, \dots, n\}$\,.
\end{lemma*}

The proof of this ``theorem on the generator" (of rings of bounded self-adjoint operators) basically follows from the spectral theorem, see p. 312 in \cite{AG1981}. It goes back to Ref. 93 in that book and is mentioned already in Sec.10 of Ch.II of \cite{vN2018}. For $\dim\H <\infty$ the proof is simple and can be found in § 84 of \cite{Hal1987}.

\paragraph{Example:} Derivation of \ref{folg_d} from \ref{folg_a} to \ref{folg_c}:

For $A=E=E^2$\;is\;$\Theta(E)=\Theta(E^2)\overset{\ref{folg_b}}{=} \Theta(E)\Theta(E) = \big(\Theta(E)\big)^2$, hence $\Theta(E)(\omega) \in \{0,1\}$ for all $\omega\in\Omega$. The claim for a general $A$ then follows from its spectral resolution combined with \ref{folg_a} and \ref{folg_c}.

\begin{proof} of the B--KS theorem (following \cite{Ho2001})

It is enough to suppose\; $3\leq\dim\H\ <\infty$. The proof is by contradiction. Let us assume that a mapping $\Theta$ with the two properties exists and consider $\mu(E)\ceq\Theta(E)(\omega)$, at a \emph{fixed} $\omega\in\Omega$, for any projection $E$ on $\H$. By the above consequences \ref{folg_a} and \ref{folg_b} it follows that $\mu$ is a probability measure on $\mathcal{P}(\H)$ that only takes the values\, $0$ or $1$, see the above example. On the other hand, by {\scshape Gleason}'s theorem, see Section 4, there is a state $W$ on $\H$ such that $\mu(E)=\tr(WE)$ for all $E$. However, for some $E\notin\{\mathbb{0},\uu\}$  the last trace is an element of the \emph{open} unit interval $]0,1[$. Contradiction!
\qed
\end{proof}

\paragraph{Remark}
The formulation of the original B--KS theorem is somewhat involved and the proof given here, though rather short and elegant, is not elementary because it is based on the above lemma and on {\scshape Gleason}'s theorem. Moreover, the contextuality does not show up explicitly.\\

Fortunately, there is a simple version of the theorem with a proof for $\dim\H\geq 4$, which is spectacularly simple, direct, and transparent with regard to contextualiy. It is due to {\scshape Mermin} \cite{Mer1993}, see also the review \cite{BCGKL2022}. It should find its way into the textbooks.  
 
\subsection*{{\scshape Mermin}'s version of the B--KS theorem}

\boxedtext{\parbox{0.974\textwidth}{If $\dim \H \geq 3$, then there is always a set $\{A_1, A_2, \dots, A_n\}$ of finitely many self-adjoint operators acting on $\H$, for which it is \emph{impossible} to assign to it a set $\{\alpha_1, \alpha_2,\dots, \alpha_n\}$ of corresponding real (eigen)values such that all functional relations between the \emph{commuting} operators among $\{A_1, A_2, \dots, A_n\}$ also hold for the assigned values.}}

\begin{proof} (for $\dim \H=4$, following \cite{Mer1993})

\begin{itemize}
 \item Let me recall the two-spin system from Sections 3.2 and 3.3 with the total {\scshape Hilbert} space
  \[  \H \cong \C^2 \otimes \C^2 \quad\quad (\dim \H =\dim\C^2 \cdot \dim\C^ 2= 2\cdot 2 = 4) \]

\item Each observable $A=A^*$ on $\H$ is a real linear combination of the 16 product observables 
   \[ \uu \otimes \uu,\; S^{\gamma} \otimes \uu,\; \uu \otimes S^{\delta},\; S^{\gamma} \otimes S^{\delta} \quad\quad (\gamma,\delta \in \{x,y,z\})\,.
  \footnote {To simplify the notation, $\uu$ stands here and in the next section, unlike in Section~3, for the unit operator $\uu_1$ resp. $\uu_2$, when it appears as a left resp. right tensor factor, that is, when it refers to the {\scshape Hilbert} space of the first resp. second spin. This reference also applies to the three spin components $S^{x}$, $S^{y}$, and $S^{z}\,.$}\]
  \item Especially, let me consider nine different product observables, arranged as entries 
   of a ${3\times 3}$ matrix (or ``magic square'', \cite{BCGKL2022}):
   \medskip
   \begin{center}
    \begin{tabular}{ccc|c}
     $S^{x} \otimes \uu$ &$\uu\otimes S^{x}$ &$S^{x} \otimes S^{x}$ &$    	\uu\otimes \uu$\\
     $\uu \otimes S^{y}$ &$S^{y}\otimes \uu$ &$S^{y} \otimes S^{y}$ &$	\uu\otimes \uu$\\
     $S^{x} \otimes S^{y}$ &$S^{y} \otimes S^{x}$ &$S^{z} \otimes
     	S^{z}$ &$\uu \otimes \uu$\\
     \hline
     $\uu \otimes \uu$ &$\uu\otimes \uu$ &$-\uu \otimes \uu$
    \end{tabular}
   \end{center}
 \item Facts:
  \begin{enumerate}[label=\alph*)]
  
   \item\label{beob_a} Each row and each column contains only pairwise \emph{commuting} observables. In this sense each row and each column constitutes a ``physical context'', that is, a set of observables whose values could, in principle, be jointly measured (and precisely determined in a common eigenstate). Therefore each observable can be considered to belong to two different contexts.
   \smallskip
    \item\label{beob_b} The (usual) product from the three observables in the \emph{right column} gives $-\uu\otimes \uu$. In contrast, the products from the respective three observables in the two other columns gives $\uu\otimes \uu$. The latter also holds for the products in each of the three rows.
    \smallskip
   \item\label{beob_c} By the assumption of \emph{functional consistency} the identities of \ref{beob_b} imply the assigned (eigen)values $-1$ resp. $1$.
Consequently, the \emph{product of all nine values} is then $-1$ by the column identities, whereas it is $1$ by the row identities. Contradiction!
 \qed
 \end{enumerate}
 \end{itemize}
 \end{proof}
 
 \paragraph{Remark}
 
In the case $\dim \H \geq 5$ it suffices to consider a 4-dimensional subspace.
Unfortunately, the proof in the case $\dim \H = 3$ remains more complicated \cite{Per2002}.

\section{Correlation Inequalities of {\scshape Bell} and Others}

\paragraph{Attention please:}\,

\medskip
In the following theorem, unlike in previous sections, the commutator of two operators $X$ and $Y$ is defined as $[X,Y]\ceq XY-YX$, that is, without the factor $\ii/\hbar$.

\medskip
\boxedtext{\begin{minipage}{\textwidth-11pt}
\begin{theorem*} ({\scshape Bell} 1964, {\scshape Clauser}--{\scshape Horne}--{\scshape Shimony}--{\scshape Holt} 1969; {\scshape Tsirelson} 1980)

\medskip
Let $A, B, C, D$ be four self-adjoint operators on an at least 4-dimensional {\scshape Hilbert} space $\H$ with the properties\, $A^2=B^2=C^2=D^2= \uu$\, and\, $[A, C]=[A,D]=[B,C]=[B,D]=\mathbb{0}$.
 Moreover, consider the {\scshape Bell}-type operator $K:= A (C+D)+ B(C-D)$ \,and\,the expectation functional  $\langle\boldsymbol{\cdot} \rangle \ceq \tr(W \boldsymbol{\cdot})$ induced by a state\, $W$\, on $\H$.\medskip

Then one has:
\begin{enumerate}[label=\roman*)]

\item\label{satz_i} $K^*=K$\; and\; $\mathbb{0}\leq K^2 = 4\,\uu - [A, B][C,D]$
 \smallskip 
  \item\label{satz_ii} $\langle K^2\rangle\leq 8 $\qquad ({\scshape Tsirelson} inequality)
 \smallskip 
  \item\label{satz_iii} If $\langle [A, B][C, D]\rangle\geq 0$\,, then \,$\langle K^2\rangle\leq 4$\qquad ({\scshape B--CHSH} inequality)\,.
 \end{enumerate}
 \end{theorem*} 
\end{minipage}}

\begin{proof}
\begin{enumerate}[label=\roman*)] \item By $(X+Y)^*=X^*+Y^*$\,,
 $(XY)^*=Y^*X^* $\,, and\, explicit squaring.
 \smallskip
\item The subadditivity (or triangle inequality) and submultiplicativity (or multiplicative inequality) of the uniform operator norm, see Explanation~\ref{skalarprodukt} in Section 2.2, gives:\newline\; $\langle K^2\rangle \leq\big\|K^2\big\| \overset{\ref{satz_i}} {\leq} \big\|4\,\uu\big\| + \big\|[A, B]\big\|\big\|[C, D]\big\|\leq 4 + (1+1)(1+1) = 8\,.$
\smallskip
\item Obviously:\; $\langle K^2\rangle\overset{\ref{satz_i}}=4-\langle [A, B][C, D]\rangle\leq 4\,.$
\qed
\end{enumerate}
\end{proof}

\paragraph{Remarks:}

\begin{itemize}
\item  By the positivity of general variances,\, \ref{satz_ii} resp. \ref{satz_iii}  implies\; $|\langle K\rangle| \leq 2 \sqrt{2}$\; resp. $|\langle K\rangle| \leq 2\,.$
\item  For the above simple proof and proofs of extensions see \cite{T/C1980,La1987,Stre2007}. One should note the generality of the above result. Neither ``stochastic independence'' nor ``locality'' is assumed, whatever the latter may mean. The interpretation of the theorem is as simple as its proof. It merely states that non-commutativity can make a difference in obtaining upper bounds on quantum expectation values. In the present case, the two (anti-self-adjoint) commutators $[A, B]$ and $[C, D]$ commute with each other, which implies that they have a common eigenbasis in $\H$. Nevertheless, it is not obvious how to measure the observable $K^2$ or even $K$ itself in realizable experiments \cite{Per2002,LeB2006}, but see \cite{RZBB1994}. 
\smallskip
\item 
Calling $\langle K^2\rangle\leq 4$ the ``{\scshape B--CHSH} inequality'' is justified, because it also holds, when $A, B, C, D$ are interpreted as four [commuting] \emph{classical} random variables taking values $\pm 1$ and the angular brackets $\braket{\boldsymbol{\cdot}}$ as the expectation with respect to an (unknown) \emph{arbitrary} joint probability measure/distribution for these values, as in the \HV-model of {\scshape B--CHSH}\; [$K^2 = 4 $\; implies\; $K=\pm 2$\; for all $2^4 = 16$ realizations], see the second paper re-printed in \cite{Bel2004} and \cite{CHSH1969}.

\end{itemize}
\subsection*{Application: Spin correlations in the singlet state}
\begin{itemize}
\item QM system: Again composed of two spins with $\H = \H_1 \otimes \H_2$,\, $\H_1 = \H_2 \cong\C^2$
  
\item Two arbitrary spin components on $\H_1$ resp. $\H_2$ as observables on $\H$ in terms of four \emph{unit} vectors $\vv{a}, \vv{b}, \vv{c}, \vv{d} \in \R^3$\,:
\[\boxedeqn{
   A\ceq\vv{a}\boldsymbol{\cdot} \vv{S} \otimes \uu,\; B\ceq\vv{b}\boldsymbol{\cdot}\vv{ S} \otimes \uu, \; C\ceq\uu \otimes \vv{c}\boldsymbol{\cdot}\vv{S},\;  D\ceq\uu \otimes \vv{d}\boldsymbol{\cdot}\vv{S}\,.}\]
Obviously, they satisfy the assumptions of the above theorem.
  
 \item \emph{Singlet state} on $\H$ [in manifestly spin-rotation invariant form]:  \[\boxedeqn{
   W^-\ceq\frac{1}{4} \left(\uu \otimes \uu - \left(\vv{S} \otimes \uu \right)\boldsymbol{\cdot} \left(\uu \otimes \vv{S} \right) \right)}\;\, \widehat{=}\,|\Phi^-\rangle\langle\Phi^-|
  \,.\]
It is entangled  with completely mixed reduced states, see Section 3.2\,. Moreover, it satisfies
  \[(W^-)^2=W^- \quad \text{and}\quad (\vv{S}\otimes \uu + \uu \otimes \vv{S})^2 W^- =\mathbb{0}\,.\]
It is therefore a pure eigenstate of the squared total-spin vector with eigenvalue
$0=4\cdot 0(0+1)$, which justifies its name. Finally, it is the (unique) ground state of the 
{\scshape Hamilton}ian\; $H\ceq\sv\, (\vv{S} \otimes \uu)\boldsymbol{\cdot} (\uu \otimes \vv{S})\ceq\sv\sum_{\gamma\in\{x,y,z\}} S^{\gamma}\otimes S^{\gamma}$,
 coupling the two spins isotropically and anti-ferromagnetically: $HW^- =-3\sv W^-\,,\; \sv> 0$\,.
\item Combining the four observablen $A, B, C, D$\,, as defined above, to a [special] $K$ as in the theorem and computing its expectation value in the singlet state $W^-$ gives explicitly in terms of four correlation coefficients:
  \begin{align*}
 \braket{K} &= \braket{AC} + \braket{AD} + \braket{BC} - \braket{BD}\\
  &= \kappa_{A, C} + \kappa_{A, D} + \kappa_{B, C} - \kappa_{B, D}\\ 
  &=-\vv{a}\boldsymbol{\cdot} \vv{c} - \vv{a}\boldsymbol{\cdot} \vv{d} - \vv{b}   			\boldsymbol{\cdot} \vv{c} + \vv{b}\boldsymbol{\cdot} \vv{d}\,.
 \end{align*}
   
 \item If all four unit vectors  $\vv{a}, \vv{b}, \vv{c}, \vv{d}$ are \emph{coplanar} and aligned as in the subsequent figure, then one finds $\braket{K} = - \sqrt{8}$. Combining this result with the {\scshape Tsirelson} inequality gives $8=\braket{K}^2\leq\braket{K^2}\leq 8$. Hence, $W^-$ is an eigenstate of this particular $K$ with eigenvalue $-\sqrt{8}$ and yields a \boxedtext{maximum violation of the classic(al) {\scshape B--CHSH} inequality.}
\begin{figure}[h!]
\begin{center}
\psset{unit=1cm}
\psset{arrowscale=2}
\begin{pspicture}*[algebraic](-2.5,-0.3)(2.7,2.5)
\pswedge[linewidth=0.1pt,linecolor=white,fillstyle=solid,fillcolor=green!30](0,0){1.4}{0}{45}
\pswedge[linewidth=0.1pt,linecolor=white,fillstyle=solid,fillcolor=green!30](0,0){1.5}{45}{90}
\pswedge[linewidth=0.1pt,linecolor=white,fillstyle=solid,fillcolor=green!30](0,0){1.6}{90}{135}

\psarc[linecolor=green]{->}(0,0){1.4}{0}{45}
\psarc[linecolor=green]{->}(0,0){1.5}{45}{90}
\psarc[linecolor=green]{->}(0,0){1.6}{90}{135}

\psline[linewidth=0.5pt,linecolor=blue]{->}(0,0)(2,0)
\rput(2.3,0){\textcolor{blue}{$\vv{b}$}}

\psline[linewidth=0.5pt,linecolor=blue]{->}(0,0)(1.41421,1.41421)
\rput(1.71,1.71){\textcolor{blue}{$\vv{c}$}}

\psline[linewidth=0.5pt,linecolor=blue]{->}(0,0)(-1.415,1.41421)
\rput(-1.71,1.71){\textcolor{blue}{$\vv{d}$}}

\psline[linewidth=0.5pt,linecolor=blue]{->}(0,0)(0,2)
\rput(0,2.2){\textcolor{blue}{$\vv{a}$}}

\psdot(0,0)

\rput(0.9238839,0.382672){$\nicefrac{\pi}{4}$}
\rput(0.382672,0.9238839){$\nicefrac{\pi}{4}$}
\rput(-0.382672,0.9238839){$\nicefrac{\pi}{4}$}
\end{pspicture}
\caption{Four coplanar unit vectors with \\ $\protect\vv{a}\|\protect\vv{c}+\protect\vv{d}$\, and\, $\protect\vv{b}\|\protect\vv{c}-\protect \vv{d}$}
\end{center}
\end{figure}
\end{itemize}

\paragraph{Remarks:}
 \begin{itemize}
 \smallskip
 \item When one computes $\braket{K}$ not with the singlet state, but with an arbitrary
\emph{non-entangled} state, one finds $\left| \braket{K} \right| \leq 2$ for all unit vectors $\vv{a}, \vv{b},\vv{c}, \vv{d} \in \R^3$, in agreement with the classical \HV-model interpretation of the {\scshape B--CHSH} inequality \cite{Au2007}. Conversely, {\scshape Werner} has constructed \emph{entangled} (mixed) states with $\left| \braket{K} \right| \leq 2$ for all unit vectors \cite{Wer1989}. 
 \item Given these mathematical results, what is their relevance for entangled states occurring in real experiments? {\scshape Bell} was motivated by {\scshape D.~Bohm}'s 1951 spin-singlet simplification of the famous ``Gedankenexperiment'' by {\scshape Einstein}--{\scshape Podolsky}--{\scshape Rosen} in 1935 to point out that correlations between two \emph {spacelike separated} quantum systems which are in an entangled state due to interactions in the past may be stronger than is possible from any classical mechanism. A real-world, but not practical, example \cite{Per2002,LeB2006} is the (rare) leptonic decay \cite{Hus2024} of an electrically neutral pion $\pi^{0}$ (with spin $0$) into an electron $e^{-}$ and a positron $e^{+}$ flying along a straight line in opposite directions.   Their individual {\scshape Hilbert}-space factors are then naturally assigned to different locations.\\
In 1982, {\scshape A. Aspect} and coworkers set up an equivalent scenario in a basement laboratory in Orsay near Paris and found that the correlation statistics of coincidence-measurement outcomes on pairs of polarized photons agrees with the {\scshape Tsirelson} inequality, but in general not with the \emph{classical} {\scshape B--CHSH} inequality \cite{ADR1982}. Later such so-called {\scshape Bell} tests, for example by {\scshape A.  Zeilinger} and coworkers \cite{WJSWZ1998} and by other (big) groups like \cite{Chr2013} and the three ones mentioned in \cite{Mil2016}, have confirmed that the probabilistic predictions of \QM\ cannot be described by classical \HV-models which are local (or locally realistic) in the sense of {\scshape Bell}, see \cite{Bel2004,BCPSW2014,BG2016}. Some authors use this fact to claim that the \QM\ itself is genuinely non-local. I do not side with these authors, because the \QM\ is clearly local in the sense that for two dynamically uncoupled systems, even if in an entangled state, no measurement or force on the second system affects the physical behavior of the first. Nevertheless, in an entangled state the systems may be spatially correlated or ``causually non-separable'', even over huge distances. For various inspiring 
discussions, see the 7th paper re-printed in \cite{Bel2004}, {\scshape Faris} in \cite{W1995}, and \cite{Mer1993,As2007,Stre2007,HS2010,Wis2014a,Wis2014b,BG2016,Pe2017,Gr2020,Cab2021,BCGKL2022,Wer2023}. It now seems that \emph{contextuality} is largely responsible for the structural differences between classical and quantum \cite{DF2022}.
\smallskip 
\item To summarize, quantum entanglement and quantum contextuality are genuinely non-classical concepts. These days they have become key resources for quantum information, communication, and computation. Trying to describe them by classical concepts is –  quoting {\scshape R.F. Streater} – like trying to model non-{\scshape Euclid}ean geometry by using figures of a strange shape in {\scshape Euclid}ean space. It can't be done.

   \end{itemize}
I'll leave the last (sarcastic!) word to Mark {\scshape Twain}\; \\(in his memoir \textit{Life on the Mississippi}, published 1883):\\

\textit{``There is something fascinating about science. One gets such wholesale returns of\\ conjecture out of such a trifling investment of fact.''}

\section{Bibliography}

\renewcommand\refname{Selected popular books}

\renewcommand\refname{Selected textbooks and monographs}

\renewcommand\refname{Articles}

\end{document}